\documentclass[letterpaper,nofootinbib,showpac,onecolumn,eqsecnum,superscriptaddress,floatfix]{revtex4}

\usepackage[linktocpage]{hyperref}
\usepackage{xcolor}
\usepackage{tabularx}
\usepackage{soul}
\usepackage{amstext}
\usepackage{fancyhdr}
\usepackage{verbatim}
\usepackage{mathrsfs}
\usepackage{amsfonts}
\usepackage{epsfig}
\usepackage{epstopdf}
\usepackage{hyperref}
\usepackage{amssymb}
\usepackage{amsmath}
\usepackage{latexsym}
\usepackage{epsfig}
\usepackage{graphicx}
\usepackage{color}

\newcommand{\be}{\begin{equation}}
\newcommand{\ee}{\end{equation}}
\newcommand{\ba}{\begin{eqnarray}}
\newcommand{\ea}{\end{eqnarray}}

\begin{document}

\title{A search for distinctive footprints of compact binary coalescence within alternatives theories of gravity}

\author{Alejandro Casallas-Lagos}  
\email[]{alejandroc.lagos@alumnos.udg.mx}

\author{Claudia Moreno} 
\email[]{Correspondig author. E-mail: claudia.moreno@cucei.udg.mx} 

\affiliation{Departamento de F\'isica,
Centro Universitario de Ciencias Exactas e Ingenier\'ias, Universidad de Guadalajara\\
Av. Revoluci\'on 1500, Colonia Ol\'impica C.P. 44430, Guadalajara, Jalisco, M\'exico}

\author{Javier M. Antelis} 
\email[]{mauricio.antelis@tec.mx} 
\affiliation{Tecnol\'ogico de Monterrey, Escuela de Ingenier\'ia y Ciencias\\
Av. Eugenio Garza Sada 2501 Sur, Colonia Tecnol\'ogico \\
Monterrey, N.L., 64849, México}

\author{Rafael Hern\'andez-Jim\'enez} 
\email[]{rafaelhernandezjmz@gmail.com} 

\affiliation{Departamento de F\'isica,
Centro Universitario de Ciencias Exactas e Ingenier\'ia, Universidad de Guadalajara\\
Av. Revoluci\'on 1500, Colonia Ol\'impica C.P. 44430, Guadalajara, Jalisco, M\'exico}


\date{\today}

%
\bigskip
\begin{abstract}
In this review we examine the amplitude intensity associated to tensorial and non-tensorial polarization modes generated by binary systems at their inspiral stage, within the alternative theories of gravity of Brans Dicke, Rosen, and Lightman Lee. This study is performed without making an explicit use of the Transverse Traceless gauge of the General Relativity approach, and at the Newtonian limit. Consequently such additional polarization modes appear (non-tensorial) due to additional degrees of freedom in modified theories of gravitation. We model and compare the different polarization modes and strain signals for each scheme varying the sky location. Our analysis allows us to identify the locations where these modes are more intense, and, therefore susceptible to being identified for the future interferometer detector network. This gives rise to a framework in which the amplitude and the intensity of all polarization modes of general relativity and alternative hypotheses can be compared.
\end{abstract}


\maketitle


\section{INTRODUCTION}
\label{sec1}

Gravitational Waves (GWs) from binary black holes (BBH) and binary neutron stars (BNS) \cite{Abbott_2019, abbott2020gwtc2} observed by the advanced LIGO (Laser Interferometer Gravitational-Wave Observatory)~\cite{TheLIGOScientific:2014jea}, VIRGO~\cite{TheVirgo:2014hva} and KAGRA~\cite{Aso:2013eba} interferometers have launched a new era of groundbreaking discoveries. Hence, this enables us to understand novel astrophysical events associated with distant and complex sources, including stellar explosions known as supernovae whose study has founded the era of multi-messenger astronomy \cite{1972ARA&A..10..335P, Alp_2018}. The first experimental detection carried out by the LIGO Collaboration~\cite{0264-9381-32-7-074001} in 2015 was the event GW150914 \cite{PhysRevLett.116.221101}; and since then, there have been around 100 confirmed findings, most of them were generated by black hole binary systems. On the one hand, Einstein in 1916 formulated the existence of GWs, by means of his linearized version of General Relativity (GR) \cite{Einstein15b}, and when imposing the Transverse Traceless (TT) gauge uniquely two tensor polarizations are yielded: $h_+$ and $h_\times$. This scheme precisely describes how GWs propagate in space over time in idealized systems. On the other hand, alternative theories of gravity predict the existence of six polarizations: two tensorial and four non-tensorial: two vectorial $h_x$, $h_y$ and two scalar $h_b$, $h_l$. Indeed, in 1973 Eardley \cite{Eardley:1973br} defined a procedure to express these six degrees of freedom by including the Newman-Penrose (NP) coefficients \cite{Penrose:1986ca} as the foundation of such analysis. He found an effective way to describe the vector, scalar and tensorial modes without making explicit use of the TT gauge of the GR approach \cite{Nishizawa:2009bf}. Moreover, recent studies of non-tensorial polarizations within alternative theories of gravity have increased due to the possibility of them being discovered by an enhanced GWs detector network LIGO/VIRGO/KAGRA  \cite{TheLIGOScientific:2014jea, TheVirgo:2014hva, Aso:2013eba, Hagihara_2019, Hagihara:2018azu},
thus opening the prospect of investigating potential deviations of GWs propagation from a GR description~\cite{Hagihara_2020, Hilborn:2021}.

Several investigations, in alternative theories of gravity, have discussed the existence of non-tensorial polarization from BNS, binary systems of pulsars, and supernovae, as well as stochastic sources \cite{LIGOScientific:2017ycc, Isi:2017equ, Rosca-Mead:2020ehn, Nishizawa:2009bf}. For instance, its has been shown that magnetic fields are in fact present at the merger of a BNS, therefore non-tensorial modes persist. Hence implying the existence of perturbed electromagnetic waves; thereby, vector polarizations might play an important role~\cite{Jimenez_2009, Wen:2018tzl}. As well as the scalar type coming from the presence of the accretion of a matter disc around a black hole (BH) \cite{Nunez:2011ej}. Indeed, non-tensorial polarizations prevail in many theories of modified gravity, such as scalar-tensor ones \cite{Takeda:2018uai, Chatziioannou:2012rf, Thorne:1973zz, Will:2018bme}.

Without imposing the TT gauge, we work in the inspiral phase of a compact binary coalescence at the Newtonian limit \cite{Antelis:2016icm}. Then we make a comparison of GWs generation among distinct alternative theories of gravity within this limit. We review and use the waveforms generated by a quadrupolar formulation of a BBH using the Brans-Dicke \cite{Will:1989sk, Will:1994fb}, Rosen \cite{Rosen:1971qp, ROSEN1974455}, and Lightman–Lee \cite{PhysRevD.8.3293} theories. We analyze the polarization orientation angles at the position of the sky of the binary system: declination and ascension. We select the values of these parameters where the amplitudes of a GW strain signal greatly intensifies, since this particular choice yields a better feasibility to determine if future network detectors could indeed recognize non-tensorial modes~\cite{Hilborn:2021}. Then, we determine and compare their distinct amplitudes, phases, and sky localization. As well as the changes observed in their polarizations. With the aim to analyze any possible GWs detection, we reconstruct the strain signal for the non-relativistic model and for some alternative theories, thus comparing the magnitude of the signals and polarization modes. And we utilize the astrophysical parameters obtained from the GW150914 event~\cite{PhysRevLett.116.221101}, such as the mass and the sky position.

This paper is organized as follows. In section II we obtain the NP null scalars in terms of the Riemann tensor, which in turn allows us to compute the six polarization modes. In section III, we derive the antenna patterns related to tensorial and non-tensorial signals. In section IV, we study the properties of the GWs polarization modes generated for relativistic and non-relativistic models within alternative theories of gravity. Finally, in section V we present conclusions of this work. The signature of the metric used in this article is ($-,\,+,\,+,\,+$).

\section{Polarization modes of null Gravitational Waves}
\label{Newton}

One of the main properties of GWs is given by the radiation-matter interaction which describes how matter responds in presence of GWs in a region of the spacetime. The geometrical orientation of a set of masses induced by the GWs is known as \emph{polarization modes} \cite{Eardley:1973br}, and in fact GR predicts two different states of polarization called plus ($+$) and cross ($\times$). In order to describe the effects of GWs on matter \cite{1972ARA&A..10..335P, poisson2014gravity}, polarization modes are measured using the geodesic equation deviation and the Riemann tensor~\cite{Wald:1984rg}. Moreover, the gravitational force is given as a scalar potential $U(x_j)$ through the equation:
\begin{equation}\label{E:Campo gravitacional}
F_j=m a_j = m \frac{d^2 x_j}{dt^2}=- \nabla U \,,
\end{equation}
where $a_j$ is the parameter of acceleration, and $U=Gm/x_j$ is the gravitational potential. Latin indices express spacial coordinates, and run from 1 to 3. Note that the physical properties induced by the action of GWs lead to the polarization states, and they can be interpreted classically by means of the geodesic deviation equation associated to a pair of masses at vacuum, where the acceleration of the masses are expressed by the formula:
\begin{eqnarray}
a_j=- R_{jtkt} x_k \,, 
\label{Riemanntensor}
\end{eqnarray}
being $R_{jtkt}$ the components of the Riemann tensor, $x_k$ represents the local coordinates associated to the mass, and $c$ is the speed of light. This equation measures the deviation of the geodesic path of two massive particles in the spacetime, and it can be seen as a first approach to the GWs-matter interaction phenomena. Indeed, this is equivalent as the work done by Einstein in 1916 \cite{Einstein15b}, where he linearized his field equations. In next section we will explain the relevance of the Riemann tensor to analyze the polarization modes. 

\subsection{Null gravitational waves in the Newman Penrose formalism}
 
The Newman-Penrose formalism \cite{Newman62a} has been used successfully, since its mathematical description of GWs constitutes a straightforward language to characterize null vectors, to compute tensorial operations in GR, and to describe the polarization states of gravitational radiation. In his 1963's original paper Ezra Newman and Roger Penrose presented a set of nineteen equations that relates a group of numbers called \emph{Newman-Penrose coefficients}, which main objective was to provide an alternative way to describe the curvature of spacetime. The Newman Penrose formalism takes as starting point a 2-dimensional complex vector space $V$, this serves as foundation to the basic objects of the theory known as spinors \cite{Penrose:1986ca}. Following the classical notation \cite{Eardley:1973br}, we consider
a null tetrad basis $(k_\mu,  l_\mu, m_\mu,  \bar m_\mu)$, being $k_{\mu}$ and  $l_{\mu}$ real null vectors, and $m_{\mu}$ a complex null vector and its complex conjugate $\bar m_{\mu}$. Then, the Minkowsky spacetime metric in terms of tetrads becomes:
\begin{eqnarray}
\eta_{\mu \nu}&=&-2l_{(\mu}n_{\nu)}+2m_{(\mu}\bar{m}_{\nu)} \,,
\end{eqnarray}
where Greek indices run from 0 to 3. The polarization components (real or complex) are encoded into the Riemann tensors, and they can also be described through the components of the Weyl and Ricci tensors in the Newman Penrose formalism \cite{Chandrasekhar:1985kt}. Thus, these components help us to define the structural form of the GWs propagation. In table \ref{scalarcomponents}, we indicate some of the main features associated to GWs: its type given by the Petrov classification scheme; the helicity through the $E(2)$ classification~\cite{Will:2018bme, Eardley:1973br, poisson2014gravity}; its components provided by the Newman Penrose coefficients, as well as their equivalence to the Weyl and Ricci tensors. In order to analyze the asymptotic behaviour of their components, the {\it Peeling theorem} \cite{Wald:1984rg, Teukolsky:1972my} states that its decay for each type is proportional to $\Psi_n = O (r^{-5+n})$ for $n=0,1..4$, and $\varphi_n = O (r^{-3+n})$ for $n=0,1,2$ ($\Phi_{nn} \equiv \varphi_n \varphi_n$). Importantly, table \ref{scalarcomponents} does not included all Weyl and Ricci polarization modes, we only cover the necessary ones to describe GWs polarization modes relevant to our study.
\begin{table}[ht]
\begin{center}
\begin{tabular}{| c | c | c |c|  }
\hline
\multicolumn{4}{ |c| }{Weyl polarization modes} \\ \hline
TYPE & HELICITY & COMPONENT & DECAY   \\ \hline \hline
Ingoing \, gravitational \, wave & $s=+2$ & $\Psi_{0} \equiv C_{kmkm}$ & $1/r^5$ \\ \hline 
Ingoing \,electromagnetic \, wave & $s=+1$ & $\Psi_{1} \equiv C_{klkm}=C_{\bar{m}mkm}$ & $1/r^4$\\ \hline
Scalar \, wave & $s=0$ & $\Psi_{2} \equiv C_{km\bar{m}l}=\frac{1}{2}(C_{klkl}+C_{kl\bar{m}m})
        =\frac{1}{2}(C_{\bar{m}m\bar{m}m}+C_{kl\bar{m}m})$ & $1/r^3$ \\  \hline
Outgoing \, electromagnetic \, wave & $s=-1$ & $\Psi_{3} \equiv C_{kl\bar{m}l}=C_{\bar{m}m\bar{m}l}$ & $1/r^2$ \\ \hline
Outgoing \, gravitational \, wave & $s=-2$ & $\Psi_{4} \equiv C_{\bar{m}l\bar{m}l }$ & $1/r$ \\ \hline
\multicolumn{4}{ |c| }{Ricci polarization modes} \\ 
\hline
Scalar \, wave & $s=0$ & $\Phi_{22}\equiv \frac{1}{2}R_{ll}$ & $1/r^2$ \\ 
\hline
\end{tabular}
\caption{Weyl and Ricci polarization modes for outgoing and incoming waves. The sub-index $k, l, m, \bar{m}$ are the tetrad projection over the tensor components, for example $\Psi_0 \equiv C_{k m k m} = C_{\mu \nu \alpha \beta} k^\mu m^\nu k^\alpha m^\beta$; the helicity or spin weight is described by the $E(2)$ group of symmetry; and the decay induced by the Peeling theorem is also included.}
\label{scalarcomponents}
\end{center}
\end{table}
A particular case of interest of the GWs dynamics is the \emph{null wave} propagation one~\cite{Eardley:1973br}, this corresponds to a GW propagating in the vacuum. Note that Riemann tensor components depend only of the retarded $t_{ret}=t-z$. Using the null condition, we can give a precise description of the NP scalars, which are associated to the way GWs propagate accordingly to the null wave conditions~\cite{Nishizawa:2009bf}. Then, the explicit forms of the non vanishing Riemann components under the null condition are:
\begingroup
\allowdisplaybreaks
\begin{align}
&\Psi_{0}=R_{kmkm}=0\,,\hspace{2.5cm}
\Psi_{1}=R_{klkm}-\frac{1}{2}R_{km}=0 \,,\hspace{1cm} \Psi_{2}=R_{km\bar{m}l}-\frac{1}{12}R=\frac{1}{6}R_{klkl} \,,\nonumber\\
&\Psi_{3}=R_{kl\bar{m}l}-\frac{1}{2}R_{l\bar{m}}=\frac{1}{2}R_{kl\bar{m}l} \,,
\hspace{1cm} \Psi_{4}=R_{\bar{m}l\bar{m}l}= R_{\bar{m}l\bar{m}l} \,,\hspace{1cm}
\Phi_{22}=\frac{1}{2}R_{ll}=R_{ml\bar{m}l}=R_{ml\bar{m}l} \,. 
\label{NPnull_relation}
\end{align}
\endgroup 

From eq.~\eqref{NPnull_relation} the remaining components form the driving-force matrix~\cite{Eardley:1973br}, which is written under the null-propagation condition in terms of the NP-null scalars:
\begin{align}
R_{titj}=
\begin{pmatrix}
R_{txtx} & R_{txty} & R_{txtz} \\
R_{tytx} & R_{tyty} & R_{tytz} \\
R_{tztx} & R_{tzty} & R_{tztz} \\
\end{pmatrix}
=\begin{pmatrix}
-\frac{1}{2}[{\rm Re} \,\Psi_{4} +\Phi_{22}] & \frac{1}{2}{\rm Im} \, \Psi_{4} & -2{\rm Re} \, \Psi_{3} \\
\frac{1}{2}{\rm Im } \, \Psi_{4} & \frac{1}{2}[{\rm Re}\, \Psi_{4}-\Phi_{22}] & 2{\rm Im} \, \Psi_{3} \\
-2{\rm Re} \, \Psi_{3} & 2{\rm Im} \, \Psi_{3} & -6\Psi_{2}
\end{pmatrix} \,.
\label{DFmatrix}
\end{align}

These six real degrees of the NP-null scalars correspond to the polarization amplitude can be expressed in terms of the Riemann tensor where $\Psi_4$ represent the plus and cross polarization, $\Phi_{22}$ gives the transverse breathing polarization, and $\Psi_2$ gives the longitudinal polarization, $\Psi_3$ stand for the vector-$x$ and vector-$y$ polarization.

\subsection{Riemann curvature tensor in first order perturbation}

\begin{figure}[thpb]
      \centering
      \includegraphics[scale=0.4]{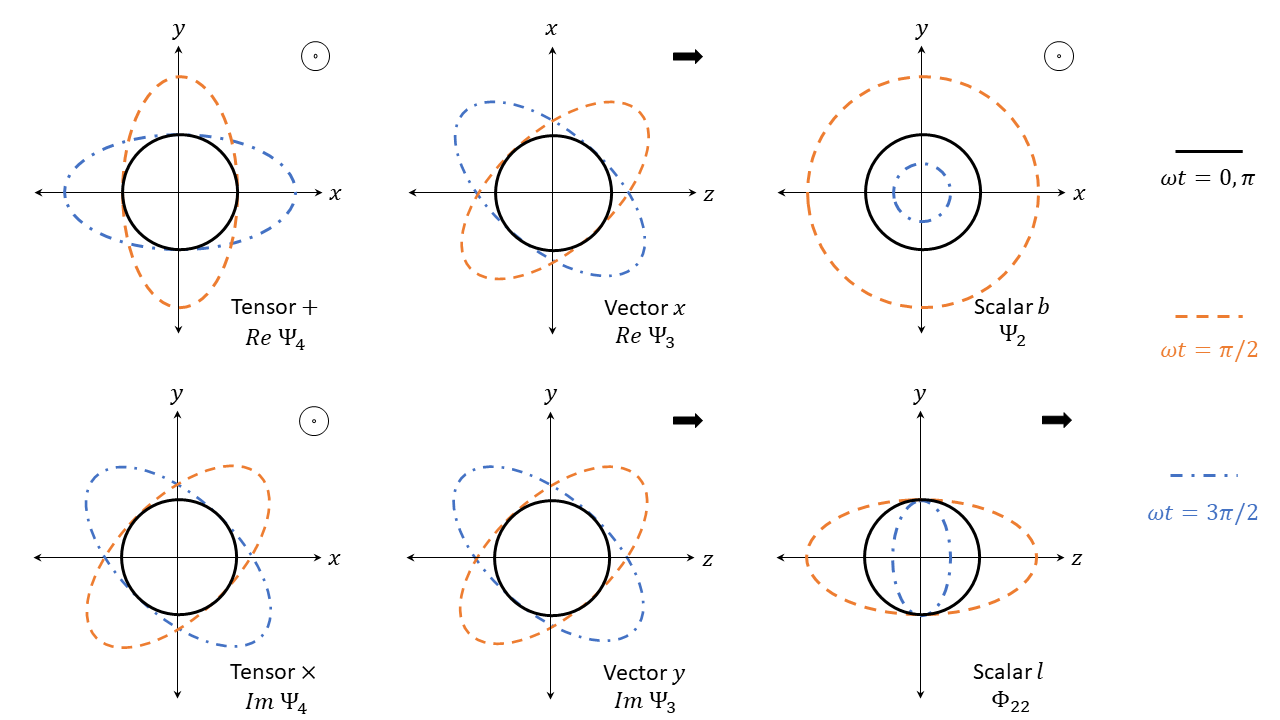}
     \caption{The six polarization modes for GW and its corresponding Newman-Penrose coefficients (see table 1). In order to describe the effect of the GW on a set of masses we consider the black arrows and the circles as the GW propagation direction; note the circle denote that the GW is pointing inwards the paper.  
     The left column contains the $+$ and $\times$ tensor modes, in the middle column the $x$ and $y$ vector modes, the right column the longitudinal $l$ and breathing $b$ scalar modes. A circle of test masses is distorted differently for each polarization propagating on the z-direction as a function of time ($\omega\, t = 0,\, \pi/2, \, \pi, 3 \,\pi/2$).}
      \label{interferometer}
  \end{figure}

In this section we will implement the linearized Einstein equations, or the weak field approximation at first order. The metric is written as $g_{\mu \nu} = \eta_{\mu \nu} + h_{\mu \nu}$ where $|h_{\mu \nu}| \ll 1$. In this scheme the Riemann tensor~\cite{dInverno:1992gxs} becomes:
\begin{eqnarray}
R^{(1)}_{\alpha \beta \gamma \delta} = \frac{1}{2} \left(h_{\beta \gamma; \alpha \delta} + h_{\alpha \delta ; \beta \gamma} - h_{\alpha \gamma; \beta \delta}- h _{\beta \delta ; \alpha \gamma}  \right) \,,
\label{NP}
\end{eqnarray}
where the superscript ${}^{(1)}$ denotes the first order of the perturbation, and ``$;$" represents the covariant derivative. For a plane GW traveling in the $z$ direction \cite{1972ARA&A..10..335P}, the perturbed tensor metric $h_{\mu \nu}=h_{\mu \nu}(t-z)$ will not depend on $x$ and $y$ coordinates. Note that having this consideration we can eliminate some derivative components of the Riemann tensor $h_{\mu \nu ;x}=h_{\mu \nu ;y}=0$. Additionally, using the Lorentz gauge condition $h^\mu_{\nu, \mu}-\frac{1}{2} h_{,\mu}=0$, we can remove the four time perturbed metric components $h_{t\mu}$. Once having the simplified perturbed Riemann components, one can obtain the Weyl and Ricci components in eq.~\eqref{DFmatrix} as well. Therefore they are~\cite{Eardley:1973br, Will:1993hxu}: 
\begin{eqnarray}
\Psi_2&=&\frac{1}{12} h_{kk,tt}=\frac{1}{12} h_{zz,tt}=-\frac{1}{6} \ddot h_l \,, \label{polarization1} \\
\Psi_3&=&\frac{1}{4} h_{k \bar m,tt}=\frac{1}{2}(h_{xz,tt}-i h_{yz,tt})= -\frac{1}{2}(\ddot h_x - i \ddot h_y)\,, \label{polarization2} \\
\Psi_4&=&\frac{1}{2} h_{\bar m \bar m,tt}=\frac{1}{2}(h_{xx,tt}-h_{yy,tt})-i h_{xy,tt}=-2(\ddot h_+ + i \ddot h_\times)\,, \label{polarization3} \\
\Phi_{22}&=&\frac{1}{2} h_{ m \bar m,tt}=\frac{1}{2}(h_{xx,tt}+ h_{yy,tt})=-2 \ddot h_b\,,
\label{polarization4}
\end{eqnarray}
where double dots represent second time derivative. Moreover, the Riemann components expressed in terms of the six real degrees of the NP-null scalars allow us to find the six polarization modes for the propagation of the GW; and these are classified as scalar: $b, l$; vectorial: $x,y$; and tensorial: $+,\times$. Furthermore, $\Psi_4$ and $\Phi_{22}$ describe transverse waves; $\Psi_2$ a longitudinal one; and $\Psi_3$ is a mixed of both type of waves. And, $\Psi_3$ and $\Psi_4$ are complex, while $\Phi_{22}$ and $\Psi_2$ are real, see Fig. \ref{interferometer}.

\section{ANTENNA PATTERN}
\label{antennapattern}

The gravitational radiation is decoded through the interaction between a GW and a set of laser-interferometric detectors placed on earth~\cite{1987MNRAS.224..131S, 1987MNRAS.226..829T}. The mathematical description of the incoming radiation is given using the perturbed metric $h_{ij}$, which contains the physical information about the GW once this is detected by the interferometer. In order to obtain a set of numerical data from the GW~\cite{Takeda:2018uai}, the metric perturbation is decomposed as a linear combination of the six polarization modes $P=\{+, \times, x, y, b, l\}$; and a set of vectors denoted by $e^{P}_{ij}({\bf \hat \Omega})=(e^+ _{ij}, e^{\times} _{ij}, e^b _{ij}, e^l _{ij}, e^x _{ij}, e^y _{ij})$, where $\bf \hat \Omega$ is the spatial orientation of the GW in the sky, and the superscript $A$ runs over the polarization modes~\cite{Nishizawa:2009bf}. To construct an analytical form of a GW, it is necessary to write down the explicit form of the polarization modes using a Cartesian coordinate system $(\hat e_x, \hat e_y, \hat e_z)$, that relates the GW to the detector in earth as follows:
\begin{equation}
\hat e_x=(1,0,0) \,, \hspace{2cm} \hat e_y=(0,1,0) \,, \hspace{2cm} \hat e_z=(0,0,1) \,,
\end{equation}
being $\hat e_z=-{\bf \hat \Omega}$ a unit vector in the direction of the GW propagation, where $\hat e_z=\hat e_x \times \hat e_y$~\cite{Nishizawa:2009bf}. However, the metric perturbation can be written in a more concise way~\cite{Takeda:2018uai}:
\begin{equation}
h_{ij}(t, {\bf \hat \Omega})=h_P(t)e^{P}_{ij}(\hat{\bf \Omega}) \,.
\end{equation}
Then, the interferometric detector has two L-shaped arms that admit a geometric representation in an orthonormal coordinate system spanned by the set:
\begin{equation}
{\bf \hat u}=(1,0,0)\,, \hspace{1cm} {\bf \hat v}=(0,1,0)\,, \hspace{1cm} {\bf \hat w}=(0,0,1) \,,
\end{equation}
therefore each arm can be considered as a Cartesian axis oriented by $\bf \hat u$ and $\bf \hat v$ directions. On the other hand, the detector responses to a signal associated to a GW, which is defined by the set $(\bf \hat u, \bf \hat v)$ through the \emph{Detector Tensor} $d$, defined by \cite{Takeda:2018uai, Nishizawa:2009bf}:
\begin{equation}\label{E:D}
d \equiv \frac{1}{2}[\bf{\hat u \otimes \hat u - \hat v \otimes  \hat v}] \,.
\end{equation}
From the geometry of the detector, eq.~(\ref{E:D}), the antenna pattern function $F_A$ is defined as: 
\begin{equation}\label{E:FA}
F_P ({\bf{\hat \Omega}}) = d:{e}^P_{ij} ({\bf \hat \Omega}) \,,
\end{equation}
here colon is a double tensor contraction operation. Then, we will introduce the antenna pattern functions, that depend of the geometry of the detector frame, $(\bf \hat u, \, \bf \hat v, \,\bf \hat w)$; as well as the representation of an incoming GW, $h_{ij}(t, {\bf \hat \Omega})$. These functions are obtained under a change of coordinates at the detector frame, so one considers rotations about an axis pointing from the source to the detector through the radiation frame~\cite{Takeda:2018uai, Nishizawa:2009bf, Hyun:2018pgn}; thereby, the explicit forms of the antenna pattern functions are~\cite{Takeda:2018uai}:
\begin{align} 
F_{b} &= -\frac{1}{2} \sin^{2} \theta \cos 2 \varphi = - F_{l} \label{Fb0} \, , \\
F_{x} &= \sin \theta \left( \cos \theta \cos 2 \varphi \cos \psi - \sin 2 \varphi \sin \psi \right) \label{Fx0} \,, \\
F_{y} &= - \sin \theta \left( \cos \theta \cos 2 \varphi \sin \psi + \sin 2 \varphi \cos \psi \right) \label{Fy0} \,, \\
F_{+} &= \frac{1}{2} \left( 1 + \cos^{2} \theta \right) \cos 2 \varphi \cos 2 \psi - \cos \theta \sin 2 \varphi \sin 2 \psi \label{Fpl0} \,, \\
F_{\times} &= -\frac{1}{2} \left( 1 + \cos^{2} \theta \right) \cos 2 \varphi \sin 2 \psi - \cos \theta \sin 2 \varphi \cos 2 \psi  \label{Fcross0} \,.
\end{align} 
Then, using eqs.~\eqref{polarization1}-\eqref{polarization4} and eqs.~\eqref{Fb0}-\eqref{Fcross0} the \emph{Strain Signal} is defined as the following configuration:
\begin{eqnarray}
h(t)&=&F_P ({\bf{\hat \Omega}}) h_P(t)e^{P}_{ij}(\hat{\bf \Omega}) \nonumber \\
&=&F_+ h_+ + F_\times h_\times+F_x h_x + F_y h_y+ F_b h_b + F_l h_l \,, \label{strain}
\end{eqnarray}
note that this equation contains information of the interaction between the GW and the detector. 

\section{TENSORIAL AND NON-TENSORIAL POLARIZATION MODES IN BINARY SYSTEMS}
\label{sec_five}

To find the GWs generated by binary systems the quadrupole formula is used, which has been obtained by using multipolar expansion in a weak gravitational field~\cite{Maggiore2007, Thorne:1980ru}:
\begin{equation}
h_{jk}=\frac{2G}{c^4 r}\frac{d^2 {\cal I}_{jk}(t-r/c)}{dt^2} \,,
\end{equation}
where $r$ is the real source-detector distance; ${\cal I}_{jk}$ is the inertia tensor, and its representation in multipolar momenta is given by the expression:
\begin{eqnarray}\label{massmoment}
{\cal I}^{jk}&=& M^{ij} = \int \rho (t, \bold x) \, x_j x_k \, d^3x \,,
\end{eqnarray}
where $\rho(t, \bold x)$ is the mass density localized at its mass center. In GR, only plus $h_+$ and cross $h_\times$ polarization modes are admitted for $M_{ij}$~\cite{Maggiore2007}.

\subsection{Binary coalescence approximation in alternative theory of GR}
\label{alternative theory}

\begin{figure}[thpb]
    \centering
    \includegraphics[width=0.6 \textwidth]{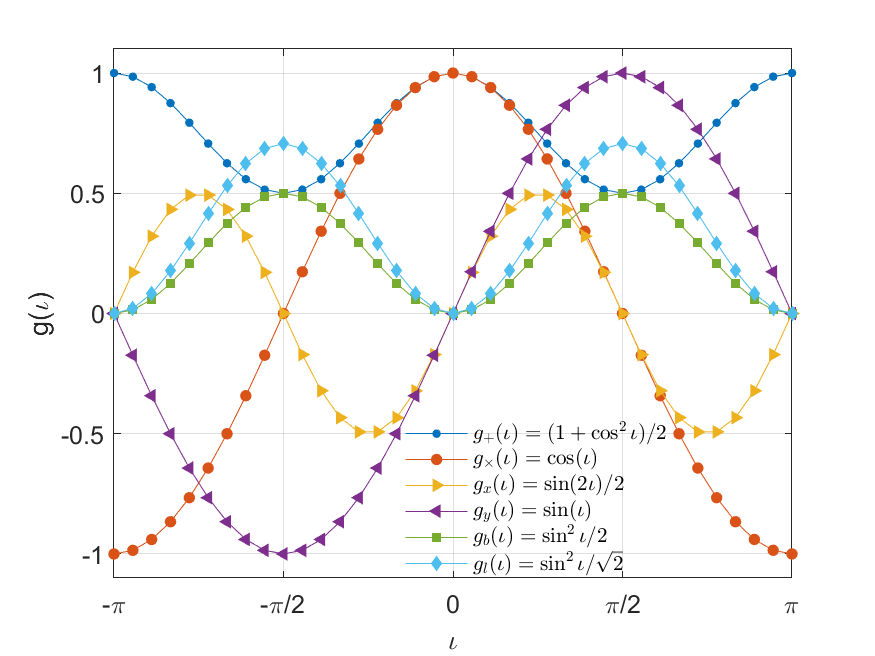}
    \caption{Variation of the functions $g_P(\iota)$, $P=\{ +, \times, x, y, b, l \}$ with respect to the inclination angle $\iota$. 
    On the one hand, for inclination angle $\iota$: $-\pi$, $0$, $\pi$, only the tensorial polarizations remain having their maximum magnitude.
    On the other hand, for $\iota$: $-\pi/2$, $\pi/2$ the polarizations $\times$ and $x$ are zero, and $y$ polarization dominates.
    Note that there are sky locations where the non-tensorial polarization modes are not detectable at all.
    }
    \label{fig:waves}
\end{figure}

In this section we will review the approximation where non-tensorial polarization modes arise from null waves, and they are characterized by the algebraic independent components of the Riemann tensor (see section III)~\cite{Takeda:2018uai, Nishizawa:2009bf}.
From alternative gravity theories~\cite{Will:1993hxu}, where the gauge symmetry is not held, emerge two tensorial polarization $h_+$, $h_\times$; and four non-tensorial polarizations of a GW: $h_x$ and $h_y$ (vectorial); $h_b$ and $h_l$ (scalar). These waves propagate perpendicular to the orbital plane in $z$ direction, and they are time dependent. To obtain the explicit expressions for the mass momenta, we choose a $(x, y, z)$ frame where the orbit lies, and the Cartesian components are expressed by: 
\begin{eqnarray}
x(t)&=& R \cos(\omega t_{ret} + \pi/2) \,, \nonumber \\
y(t)&=& R \sin(\omega t_{ret} + \pi/2) \,, \nonumber \\
z(t)&=& 0 \,,
\label{coordinates}
\end{eqnarray}
where $\pi/2$ is the value at $t=0$, $\omega$ is the angular frequency of the binary system, and $t_{ret}=t-r/c$ is the retarded time; and $R$ is the separation distance between the masses $m_1$ and $m_2$, $\mu= m_1 m_2/M$ and $M=m_1 + m_2$ are the reduced mass and the total mass, respectively. Using the coordinates eq.~\eqref{coordinates} into eq.~\eqref{massmoment}, the mass momenta are expressed in the following way:
\begin{eqnarray}
&& {M}_{11}=\mu R^2 \frac{1-\cos 2 \omega t_{ret}}{2} \,, \nonumber \\
&& {M}_{22}=\mu R^2 \frac{1+\cos 2 \omega t_{ret}}{2} \,, \nonumber \\
&& {M}_{12}=\frac{1}{2}\mu R^2 \sin 2 \omega t_{ret} \,.
\label{moment1}
\end{eqnarray}
From eq.~(\ref{moment1}) the tensorial polarizations are
\begin{eqnarray}
h_+(t)&=&\frac{\kappa}{r}  \left(\frac{1+\cos^2 \iota}{2}\right)\cos(2 \omega t_{ret}+2\phi_0) \,, \label{hplus_m}\\
h_\times(t)&=&\frac{\kappa}{r} \cos \iota \sin(2 \omega t_{ret}+2\phi_0) \,, \label{ht_m}
\end{eqnarray}
and the non-tensorial polarization are
\begin{eqnarray}
h_x(t)&=&\frac{\kappa}{r} \frac{\sin 2 \iota}{2} \cos(2 \omega t_{ret}+2\phi_0) \,, \label{hx_m} \\
h_y(t)&=&\frac{\kappa}{r} \sin \iota \sin(2 \omega t_{ret}+2\phi_0), \label{hy_m} \\
h_b(t)&=&\frac{\kappa}{r} \frac{\sin^2 \iota}{2} \cos(2 \omega t_{ret}+2\phi_0) \,, \label{hb_m} \\
h_l(t)&=&\frac{\kappa}{r} \frac{\sin^2 \iota}{\sqrt{2}} \cos(2 \omega t_{ret}+2\phi_0) \,, \label{hl_m}
\end{eqnarray}
where $\kappa = 4G\mu \omega^2 R^2/c^4$ is the signal amplitude, $r$ is the distance between the source and the observation point, $\iota$ is the inclination angle, which is the angle between the source angular momentum axis and the observer line of sight, of the binary system, and $\phi_0$ is the orbital phase. In order to obtain the $\sin$ and $\cos$ functions, in terms of the time of the observation, their argument are simplify by $2\omega t_{ret}+2\phi_0=2\omega t-2 \omega  r /c+2\phi_0=2\omega t+2\alpha$ being $2 \alpha$ a $2\pi$ multiple ($\alpha=\phi_0-\omega r/c$); thus $\cos(2 \omega t)$ and similar for $\sin$ function. Then, to analyze the emitted GWs, the orbital frequency of the source $\omega$ is related to the orbital radius $R$ by $v^2/R=GM/R^2$, where $v=\omega R$. From Kepler's third law~\cite{Maggiore2007}, the orbital radius is related to the frequency by the formula~\cite{brown2007searching}:
\begin{equation}\label{kepler}
R=\left(\frac{GM}{\omega^2}\right)^{\frac{1}{3}} \,. 
\end{equation}
A GW orbital frequency is related to the orbital frequency by the relationship $\omega_{GW}=2 \omega$, yielding: 
\begin{equation} \label{frecuency} 
 f=2\left(\frac{\omega}{2 \pi} \right) \,.
\end{equation}
Once we have expressed the orbital radius $R=R(\omega)$ and frequency $f=f(\omega)$, the parameter $\kappa$ becomes $\kappa_{freq}= 4G \mu \left(\pi G M f \right)^{\frac{2}{3}}/c^4$~\cite{brown2007searching} and $2\omega t = 2 \pi  f t = 2 \phi(t)$. Thus, all polarization, written in terms of the chirp mass $\mathcal{M}=\mu^{3/5}M^{2/5}$, are expressed by:
\begin{eqnarray}
h_{+}(t) &=& \frac{{\cal A}(t)}{r} \frac{1 + \cos^2\iota}{2} \cos (2\phi(t)) \,, \label{hmas}
\\
h_{\times}(t) &=& \frac{{\cal A}(t)}{r}  \cos\iota  \sin (2\phi(t) ) \,, \label{hcruz}
\\
h_x (t)&=& \frac{{\cal A}(t)}{r} \frac{\sin 2 \iota}{2}  \cos (2\phi(t) ) \,, \label{hxf} \\
h_y (t)&=& \frac{{\cal A}(t)}{r} \sin \iota \sin (2\phi(t) )\,,  \label{hyf} \\
h_b (t)&=& \frac{{\cal A}(t)}{r} \frac{\sin^2 \iota}{2} \cos (2\phi(t) ) \,, \label{hbf}\\
h_l (t)&=& \frac{{\cal A}(t)}{r}  \frac{\sin^2 \iota}{\sqrt{2}} \cos(2\phi(t)) \,, \label{hlf}
\end{eqnarray}
where $\phi(t)$ is orbital phase, as well as the wave amplitude ${\cal A}(t)$:
\begin{equation}\label{amplitud}
{\cal A}(t) = 4 \left( \frac{G\mathcal{M}}{c^2} \right)^{5/3} \left( \frac{\pi f(t)}{c} \right)^{2/3} \,,
\end{equation} 
and $f(t)$ is the orbital frequency of the binary system.
From eqs.~\eqref{hmas}-\eqref{hlf} one can observe that all the polarization are functions of the parameters ${\cal A}(t)$, $\iota$, and $\phi(t)$, hence we generalize these quantities as follows:
\begin{equation}
h_P=\frac{{\cal A}(t)}{r}\,g_P (\iota) \, F_P (2 \phi(t)) \,,    
\end{equation}
where the subscript $P$ stands for any kind of polarization: $P=\{ +, \times, x, y, b, l \}$. Then, $g_P$ is a function of the angle $\iota$; therefore $g_+(\iota)$ and $g_\times(\iota)$ represent the tensorial polarizations; $g_x(\iota)$, $g_y(\iota)$, $g_b(\iota)$, and $g_l(\iota)$ are the non-tensorial modes. And $F_P=F_P(2\phi(t))$ is either a sine or cosine function. Fig.~\ref{fig:waves} illustrates how $g_P(\iota)$ modifies the amplitude of the different components. On the one hand, at inclination angles $\iota= -\pi, 0, \pi$, the functions $g_P(\iota)$ for the tensorial polarizations yield their maximum magnitude, whilst for the non-tensorial sectors they are null. This upshot indeed shows the absence of non-tensorial modes in those inclination angles. On the other hand, at $\iota=-\pi/2, \pi/2$, $g_\times(\iota)$ and $g_x(\iota)$) are zero, while $g_+(\iota)$, $g_y(\iota)$, $g_b(\iota)$, and $g_l(\iota)$) remain; and in fact, $g_y(\iota)$ produces the maximum value of all. In the rest of the inclination angles all functions $g_P(\iota)$ are non-zero and thus all polarizations exist. 

\subsubsection{Tensorial and non-tensorial strain signals of binary system}

\begin{figure}[thpb]
    \centering
    \begin{tabular}{cc}
    \includegraphics[width=0.49 \textwidth]{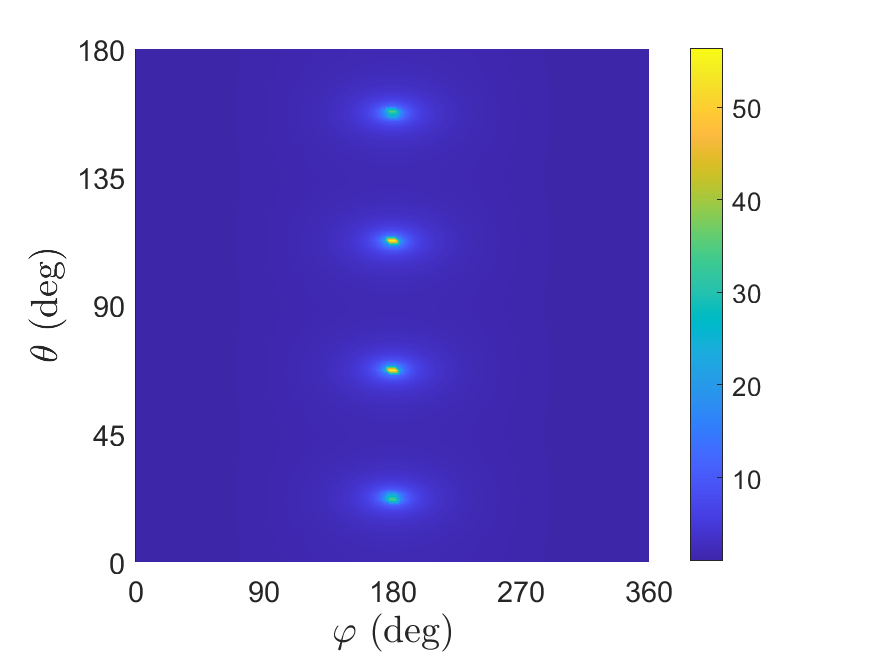}
    & 
    \includegraphics[width=0.49 \textwidth]{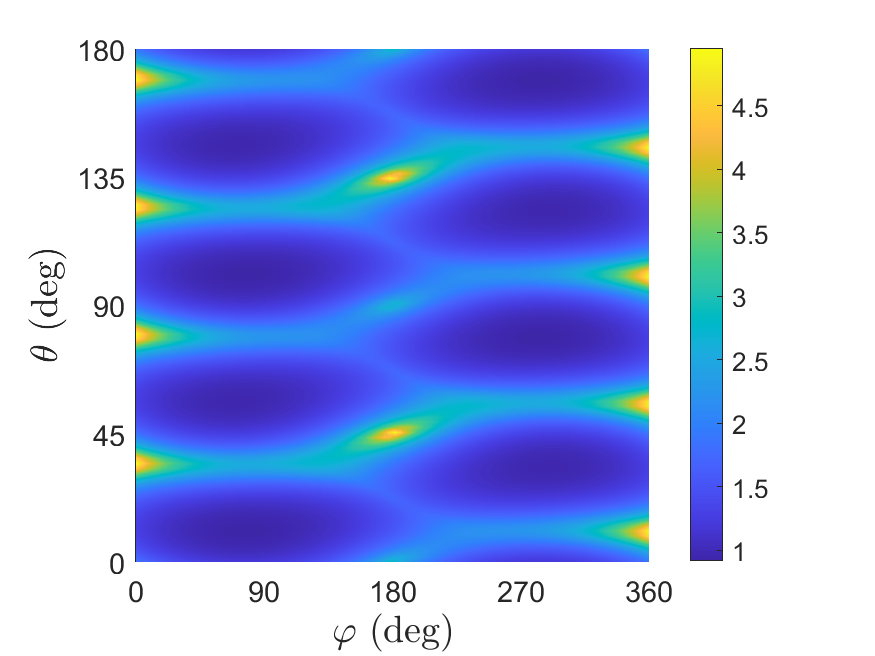}
    \\
    (a) & (b)
    \end{tabular}
    \caption{
    Dependence of the effective distance factor $1/{\cal D}_{ap}$ with respect to source orientation $\theta$ and $\varphi$ for inclination angle of (a) $\iota=0$ (optimally oriented), and (b) $\iota =\pi/3$. Note that the intensity of case (b) is smaller than (a).}
    \label{fig:effective distance}
\end{figure}

In this section we will obtain the non-relativistic GW polarizations. To obtain a strain signal additional to $h_P$, we consider the $F_P$ pattern factors for each component $P= \{+, \times, x, y, b, l\}$. Hence from eqs.~\eqref{Fb0}-\eqref{Fcross0}, it yields
\begin{equation}\label{strain}
h(t)=F_+ h_+ + F_\times h_\times+F_x h_x + F_y h_y+ F_b h_b + F_l h_l \,, 
\end{equation}
where $F_+ h_+ + F_\times h_\times$ are the tensorial components; $F_x h_x + F_y h_y$ are the vectorial and $F_b h_b + F_l h_l$ the scalar elements. Then manipulating algebraically eqs.~\eqref{Fb0}-\eqref{Fcross0}; eqs.~\eqref{hmas}-\eqref{hlf}; as well as eq.~\eqref{strain}, we obtain the expression for the strain:
\begin{eqnarray}
h(t) &=& \frac{{\cal A}(t)}{r}\left \{ \left[ F_+ \frac{1+\cos^2 \iota}{2}  + F_x \frac{\sin 2 \iota}{2} + \left(\frac{F_b}{2}+\frac{F_l}{\sqrt{2}} \right) \sin^2 \iota \right ] \cos 2\phi(t)  \nonumber \right. \\
&& \left. \hspace{1cm} + \left[ F_\times \cos \iota +F_y \sin \iota \right] \sin 2\phi(t) \right\} \,,
\end{eqnarray} 
where ${\cal A}(t)$ is the time-dependent wave amplitude \cite{whelan:2017}. We can simplify even more our formulas, so we use the trigonometric formula:
$\alpha \cos \phi + \beta \sin \phi =\gamma \cos(\phi-\Psi)$, where
$\alpha=\gamma \cos \Psi,$ $\beta =\gamma \sin \Psi$ and $\gamma=\sqrt{\alpha^2+\beta^2}$, therefore
\begin{eqnarray}
h(t) &=& \frac{{\cal A}(t)}{r}  \sqrt{ B_{NT}^2 + C_{NT}^2} \, \cos (2\phi(t) - \Psi) \,,
\end{eqnarray}
where $\Psi$ is a phase angle that quantifies the inclination and detector’s response according to:
\begin{eqnarray}
\tan \Psi &=& \frac{C_{NT}}{B_{CN}}  \,,\\
B_{NT} &=&  F_+ \frac{1+\cos^2 \iota}{2}  + F_x \frac{\sin 2 \iota}{2} + \left(\frac{F_b}{2}+\frac{F_l}{\sqrt{2}} \right) \sin^2 \iota \,, \\
C_{NT} &=& F_\times \cos \iota +F_y \sin \iota \,.
\end{eqnarray}
Finally, we introduce the effective distance between the source and the detector ${\cal D}$~\cite{Antelis2017}:
\begin{equation}\label{distance}
{\cal D} = \frac{r}{ \sqrt{ B_{NT}^2 + C_{NT}^2}} = \frac{r}{{\cal D}_{ap}} \,,
\end{equation}
where ${\cal D}_{ap}$ is the effective distance factor having a range of values: $[0, 1]$, recall that $r$ is the real source-detector distance; and $ap$ stands for all polarizations. Besides, ${\cal D}$ is equal to $r$ in its optimal orientation, i.e. when the source is located along the z-axis on the observer’s line of sight, or greater than $r$ when the source is suboptimally oriented. Thus, we get a simple form of the GW strain, as follows:
\begin{equation}\label{GWstrain} 
h(t)=\frac{{\cal A}(t)}{{\cal D}} \cos [2\phi(t) - \Psi] \,.
\end{equation}
Moreover, to illustrate the relation between $r$ and $\cal D$
Fig.~\ref{fig:effective distance} shows different values of the factor $1/{\cal D}_{ap}$ with respect to angles $\theta$ and $\varphi$. On the one hand, for $\iota=0$, the effective distance factor is close to one except for $\varphi=180$ and $\theta=22.5, 67.5, 112.5, 157.5$, and in fact at these spots ${\cal D}_{ap}$ is smaller than one, therefore the induced strain becomes stronger. On the other hand, for $\iota=\pi/3$, $1/{\cal D}_{ap} > 1$ at several orientations, which yields a smaller strain at those sky localizations. Hence, from this analysis we can infer that for some source regions, the contributions of the tensor, scalar, or vector modes are preferential at certain positions of the sky~\cite{Hagihara_2020, Hagihara_2019}.

\subsubsection{Newtonian approximation for phase and angular frequency}

When a GWs is modeled using the Newtonian limit, its signals can often be smaller than the ones generated inside the interferometers. However, performing a Newtonian expansion will allow us to improve the intensity of the gravitational trace compared to that of the detectors. Newtonian waveforms model the amplitude evolution at higher-order $v/c$ corrections of the phase evolution, hence, a more accurate solution is yielded for the motion of a binary system. In this section, we will present the tensorial and non-tensorial polarizations using the Newtonian approximation, where the phase and frequency expansion of the GW~\cite{Antelis:2016icm} are expressed by: 
\begin{eqnarray}
\phi(t) &=& - \frac{2}{5} \frac{c^3}{G M } \, \Theta(t)^{-3/8} (t_c-t) + \phi_c\,,
\label{phiTheta} \\
f(t) &=&\frac{1}{8} \frac{ c^3 }{\pi Gm} \Theta(t)^{-3/8}\,,
\end{eqnarray}
where $\Theta(t)=\frac{c^3 \eta}{5 GM }(t_c-t)$, $t_c$ is the end time of the inspiral or the coalescence time and $\phi_c$ is the value of the orbital phase at $t_c$. 

In order to exemplify the tensorial and non-tensorial polarizations at Newtonian limit, we will choose as physical parameters the observational values obtained by the GW detection of a binary black system GW150914~\cite{PhysRevLett.116.221101} which are: $m_1=36M_\odot$, $m_2=20M_\odot$, located at a distance $r=420Mpc$ and sky position $\theta=80^{\circ}$  for latitude and $\varphi=120^{\circ}$ for longitude. Importantly these parameters will be used henceforth in the paper. In addition, the waveforms are presented for two cases of inclination angle $\iota = 0,\pi/3$.

\begin{figure}[thpb]
    \centering
    \begin{tabular}{cc}
    \includegraphics[width=0.49 \textwidth]{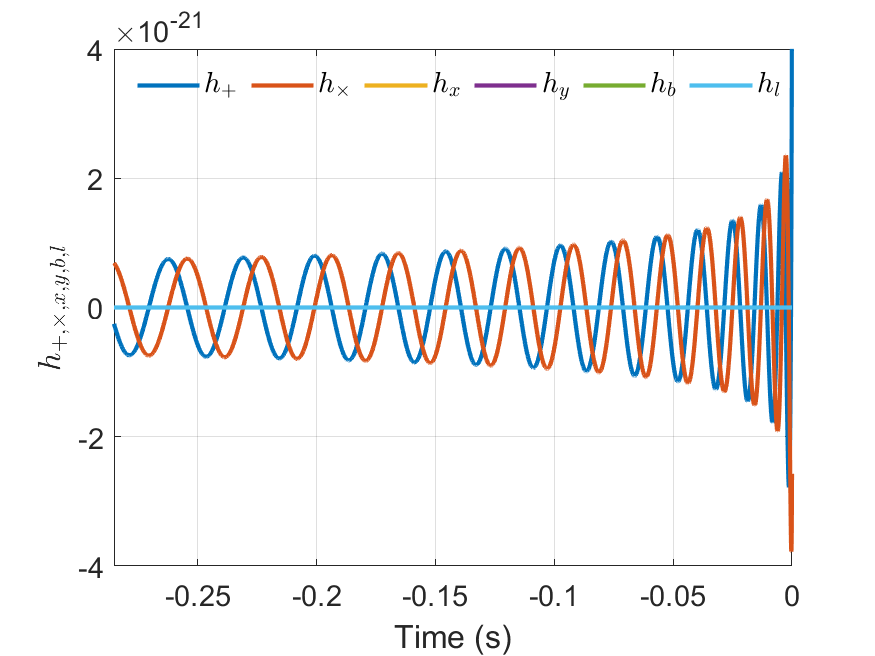}
    & 
    \includegraphics[width=0.49 \textwidth]{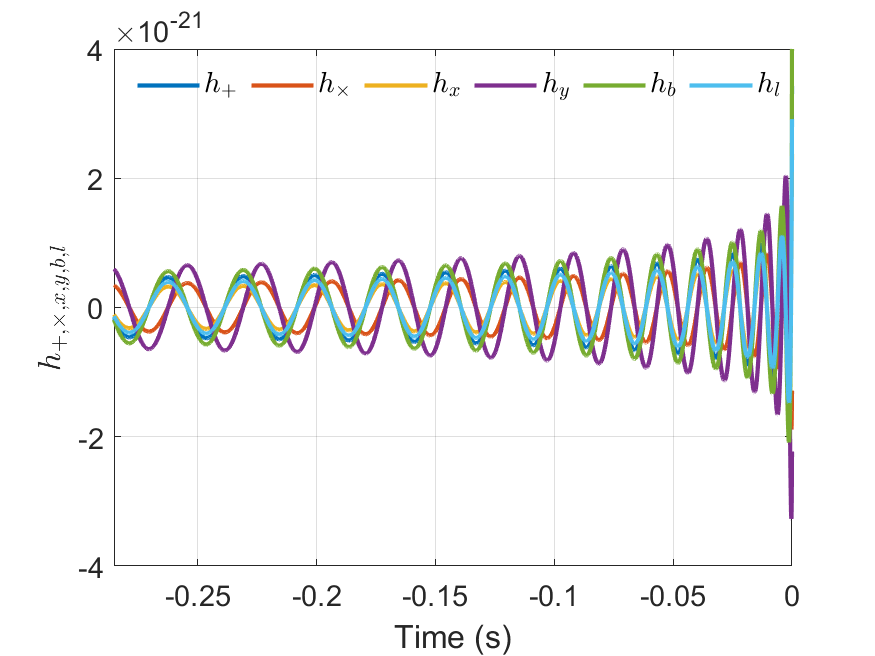}
    \\
   \includegraphics[width=0.49 \textwidth]{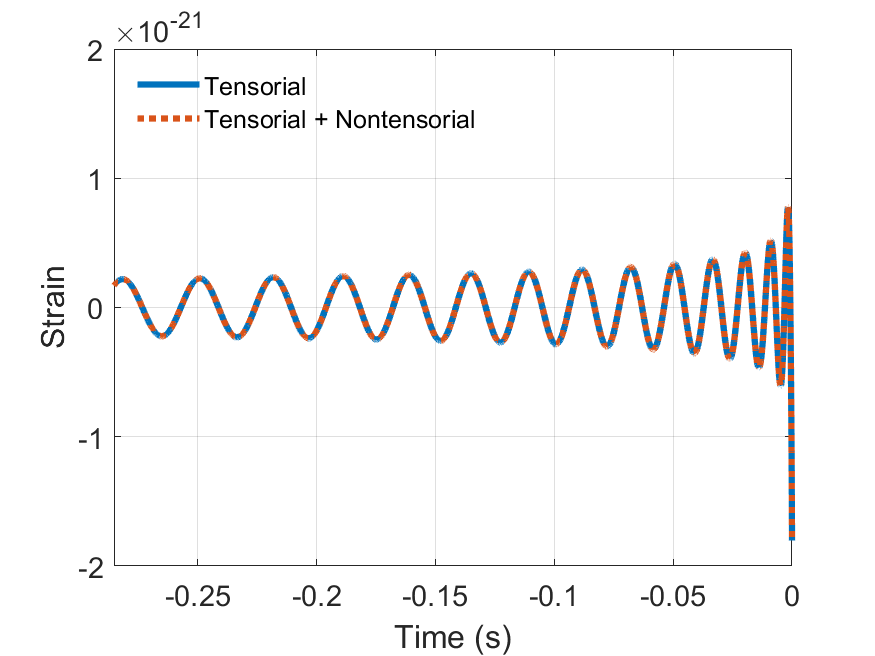}
    & 
    \includegraphics[width=0.49 \textwidth]{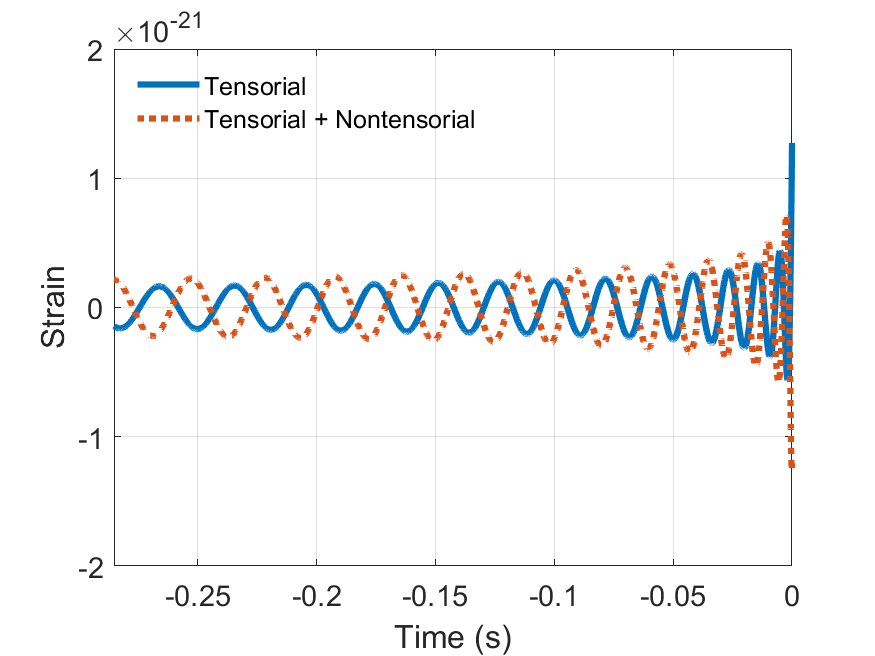}
    \\
    (a) & (b) 
    \end{tabular}
    \caption{
    Plots of tensorial and non-tensorial GW polarizations (top figures) and induced strain (bottom figures) for a binary system with parameter values from the event GW150914~\cite{PhysRevLett.116.221101}, and for inclination angles $\iota=0$ (a) and $\iota=\pi/3$ (b). Note that only tensorial modes remain when $\iota=0$, while for $\iota=\pi/3$ both polarizations are present.
    }
    \label{fig:GWpolarizations1}
\end{figure}

In Fig. \ref{fig:GWpolarizations1} we plot tensorial and non-tensorial polarizations and induced strain whit the data of GW150914 event with phase and frequency corrections at the Newtonian approximation. For $\iota=0$, only tensorial polarizations remain, therefore the induced strain is the same for both tensorial and non-tensorial modes. On the other hand, when $\iota=\pi/3$, both sectors are present, the amplitude and phase change and therefore the induced strain is different for each case. This result is according to the result showed in Fig. \ref{fig:effective distance}. 

\subsection{Binary coalescence approximation in Brans Dicke theory}

Brans Dicke theory has been extensively implemented to study the coalescence of a binary black hole system~\cite{Will:1989sk, Will:1994fb}, in particular Maggiore~\cite{Maggiore:1999wm} computed the response and the angular pattern functions of an interferometer for a scalar component of gravitational radiation. In this section, by using previous results, we analyze the dependence of the inclination and polarization angles of a binary system for the non-tensorial polarization modes. Unlike GR, Brans Dicke contains a scalar field in which gravitational interaction is affected and is inversely proportional to the gravitational constant $\cal G \approx$ 1/$\varphi_{BD}$. This scalar field produces an external influence on the structure of a compact body and it is parameterized by the dimensionless coupling constant $\omega_{BD}$.

Moreover, Brans Dicke is a scalar-tensor theory that presents an additional polarization to GR, the breathing polarization. In this section we obtain and analyze such elements for a binary compact coalescence system at the Newtonian limit, in a similar way to section~\ref{alternative theory}. In fact, we expect physical differences in GWs generation at the Newtonian limit between the relativistic and Brans Dicke schemes. Hence, to make a comparative analysis we start with the Newtonian equation of motion for a binary system in Brans Dicke theory:
\begin{equation}
    \frac{d^2 \bold x}{dt^2}=-\frac{{\cal G} m \bold x}{r^3} \,,
\end{equation}
where ${\cal G}=G-\xi (s_1+s_2 - 2 s_1 s_2)$, and $s_{n}$ is the \emph{sensitivity} of the $n$th object which measures the gravitational binding energy per unit mass; then $\xi=(2+\omega_{BD})^{-1}$ is a no-negative monopolar term, and $\omega_{BD}$ is a coupling constant. 

\subsubsection{Gravitational Waves polarization}\label{GWs-P-BD}

In Brans-Dicke theory, the metric perturbation $\bar h^{\mu \nu} \equiv \eta^{\mu \nu}-\sqrt{(-g)} g^{\mu \nu}$ is defined in terms of a covariant conserved tensor $\theta^{\mu \nu}$ and the scalar field $\varphi_{BD}$~\cite{Will:1989sk} as follows:
\be
\bar{h}^{\mu \nu}=\theta^{\mu \nu}+\frac{\delta \varphi_{BD}}{\varphi_0}\eta^{\mu \nu}\,,
\ee
where $\delta \varphi_{BD}$ is the perturbation of the scalar field $\varphi_{BD}$, and $\varphi_{0}$ is the asymptotic value of the scalar field at spatial infinity. The scalar and tensor perturbations of quasi-circular orbits~\cite{poisson2014gravity, Will:1994fb} are:
\begin{align}
\frac{\delta \varphi_{BD}}{\varphi_0}&=\frac{\mu}{r} \xi  \left\{\Gamma \left[(\hat{N}\cdot {\bold v})^2-\frac{{\cal G }M}{r^3}(\hat{N}\cdot {\bold x})^2\right]-\left({\cal G}\Gamma +2\Lambda\right)\frac{M}{r}-2 S \left(\hat{N}\cdot {\bold v}\right)\right\}\,, \\
\theta^{ij}&=\frac{2}{r} \left(1-\frac{1}{2} \xi \right) \frac{d^2}{dt^2}  \sum_{a=1,2} m_a x^i_a x^j_a =\frac{4 \mu}{r} \left(1-\frac{1}{2}\xi\right)\left({v}^i {v}^j-\frac{{\cal G}M}{r^3} {x}^i {x}^j\right)\,,
\end{align}
where $\Lambda={\cal G}(1-s_1-s_2) - \xi [(1-2s_1)s'_2+(1-2s_2)s'_1]$,  $r$ and $\hat N$ are the distance and directions unit vector of the observer, respectively. Based on the metric perturbations the GW polarization $h^{ij}$ are:
\begin{equation}
h^{ij}=\theta^{ij}-\frac{1}{2}  \left( \frac{\delta \varphi_{BD}}{\varphi_0} \right) \left(\delta^{ij}- \hat N^i \hat N^j \right) \,.
\end{equation}
Then, for quasi-circular orbits, the waveform becomes~\cite{poisson2014gravity}:
\begin{align}
\label{S-def}
h^{ij}&=\frac{2 \mu}{R} \left[2 \left(1-\frac{1}{2}\xi\right)\frac{{\cal G} M}{r}(\hat{v}^i\hat{v}^j-\hat{x}^i\hat{x}^j)\,+ \bar S (\delta^{ij}- \hat N^i \hat N^j) \right] \,, \\
\bar{S} &=\frac{\xi}{4} \left [ \frac{\Gamma {\cal G}M}{r}[(\hat{N}\cdot\hat{v})^2-(\hat{N}\cdot\hat{x})^2]-({\cal G}\Gamma +2\Lambda)\frac{M}{r}-2 S\left(\frac{{\cal G}M}{r}\right) ^{1/2} \left(\hat{N}\cdot\hat{v}\right)\right ] \,,
\end{align}
where $\xi=(2+\omega_{BD})^{-1} \sim \omega_{BD}^{-1}$ for $\omega_{BD} \gg 1$ and $\Gamma=1-2(m_1 s_2+ m_2 s_1)/M$. Indeed, one can verify that the Brans Dicke theory can be reduced to GR at the $\omega_{BD} \to \infty$ (or $\xi \to 0$) limit. In order to study a binary system within the Brans Dicke theory, we follow the algebraic steps from subsection~\ref{alternative theory}, therefore the scalar and tensorial polarization modes in terms of the orbital frequency (eq.~\eqref{kepler}) are:
\begin{align}
h_+(t)&=\left(1-\frac{1}{2}\xi \right)\dfrac{2 {\cal G}}{c^4 r} \mu \left(\pi G M f \right)^{\frac{2}{3}}\left(1+\cos^2 \iota \right) \cos\left(2\pi f t  \right) \,, \\
h_\times(t)&=\left(1-\frac{1}{2}\xi\right)\dfrac{4{\cal G}}{c^4 r} \mu \left(\pi G M f \right)^{\frac{2}{3}} \cos \iota \sin \left(2\pi f t  \right) \,, \\
h_b(t)&=\dfrac{4{\cal G}}{c^4 r} \mu \left(\pi G M f \right)^{\frac{2}{3}} \xi \left(\frac{\Gamma}{2}\sin^2 \iota \cos \left(2\pi f t  \right)- \frac{\Gamma + 2 \Lambda}{2}- (s_1-s_2) \left(\frac{r}{M}\right)^{1/2} \sin \iota \cos \left(\pi f t  \right) \right) \,, 
\end{align}
where the longitudinal and vectorial modes vanish. Finally, the all modes in terms of the chirp mass (eq.~\eqref{amplitud}) are: 
\begin{eqnarray}
h_{+}(t) &=& \left(1-\frac{1}{2}\xi \right) \frac{{\cal G}{\cal A}(t)}{r} \frac{1 + \cos^2\iota}{2} \cos (2\phi(t) ) \,, \label{hmasBD}
\\
h_{\times}(t) &=& \left(1-\frac{1}{2}\xi \right) \frac{{\cal G}{\cal A}(t)}{r} \cos\iota  \sin (2\phi(t)) \,, \label{hcruzBD}  \\
h_{b}(t) &=& \frac{{\cal G}{\cal A}(t)}{r} \, \xi \left(\frac{\Gamma}{2}\sin^2 \iota \cos \left(2\phi(t)  \right)- (s_1-s_2) \left(\frac{r}{M}\right)^{1/2} \sin \iota \cos \left(\phi(t)  \right)- \frac{\Gamma + 2 \Lambda}{2} \right) \label{hbBD} \,,
\end{eqnarray}
where the amplitude term ${\cal A}(t)$ is the same for all three expressions. However, it can be noticed distinct consequences between the non-relativistic and the above Brans-Dicke formulas, provided by the additional term $\xi$ into the amplitude factor and the constant ${\cal G}$. The breathing polarization in eq.~\eqref{hbBD} contains two harmonic terms dependent on sine and cosine functions, but the third term is a scalar that modifies the harmonicity of the wave; this last term has a direct dependence of the sensitivity constant $s_n$. Indeed, this monopolar term $h_b$ appears since the presence of scalar field $\varphi_{BD}$ in the model. To obtain a simple form of the strain function (eq.~\eqref{strain}) we utilize eqs.~\eqref{hmasBD}-\eqref{hbBD} to obtain:
\begin{eqnarray}
h(t) &=& \frac{{\cal G}{\cal A}(t)}{r}\left\{ \left[ \left(1-\frac{1}{2} \xi \right) \left(F_+ \frac{1+\cos^2 \iota}{2}  \right) + \xi \frac{F_b}{\sqrt{2}} \frac{ \Gamma}{2}\sin^2 \iota \right] \cos 2\phi(t) + \left(1-\frac{1}{2}\xi \right) F_\times \cos \iota \sin 2\phi(t)  \nonumber 
 \right. \\
&& \left.
- \xi F_b \left[(s_1-s_2) \left(\frac{r}{M}\right)^{1/2}  \sin \iota \cos \phi(t)  + \frac{(\Gamma + 2 \Lambda)}{2}\right] \right\} \,.
\end{eqnarray}
Using newly the trigonometric relation $\alpha \cos \phi + \beta \sin \phi =\gamma \cos(\phi-\Psi)$ we obtain the strain signal:
\begin{eqnarray}
h(t) &=& \frac{{\cal G}{\cal A}(t)}{r} \left [ \sqrt{B_{BD}^2+ C_{BD}^2} \cos \left(2\phi(t) - \Psi \right)
+  E_{BD}\cos (\phi(t) - \Psi)  + F_{BD} \right ] \,.
 \end{eqnarray}
where
\begin{eqnarray}
B_{BD}&=& \left(1-\frac{1}{2} \xi \right) \left(F_+ \frac{1+\cos^2 \iota}{2}  \right) + \xi \frac{F_b}{\sqrt{2}} \frac{ \Gamma}{2}\sin^2 \iota \,, \nonumber \\
C_{BD} &=& \left(1-\frac{1}{2}\xi \right) F_\times \cos \iota \,, \nonumber \\ 
E_{BD} &=& \xi F_b \left(\frac{r}{M}\right)^{1/2} (s_1-s_2) \sin \iota \,, \\
F_{BD} &=& -\xi F_b \frac{(\Gamma + 2 \Lambda)}{2} \,.
\end{eqnarray}
We define
\begin{eqnarray}\label{distance}
{\cal D}_{BD1}&=& \frac{r}{ \sqrt{ B_{BD}^2 + C_{BD}^2}}, \qquad  {\cal D}_{BD2}= \frac{r}{E_{BD}} \qquad \textrm{and} \qquad {\cal D}_{BD3}= \frac{r}{F_{BD}} \,,
\end{eqnarray}
therefore the GW strain is expressed by:
\begin{eqnarray}\label{waveformbd}
h(t) &=& {\cal A}(t) {\cal G} \left [ \frac{1}{{\cal D}_{BD1}} \, \cos (2\phi(t) - \Psi) + \frac{1}{{\cal D}_{BD2}} \, \cos (\phi(t) - \Psi)+ \frac{1}{{\cal D}_{BD3}} \right ] \,.
\end{eqnarray}
\begin{figure}[t]
    \centering
    \begin{tabular}{cc}
    \includegraphics[width=0.49 \textwidth]{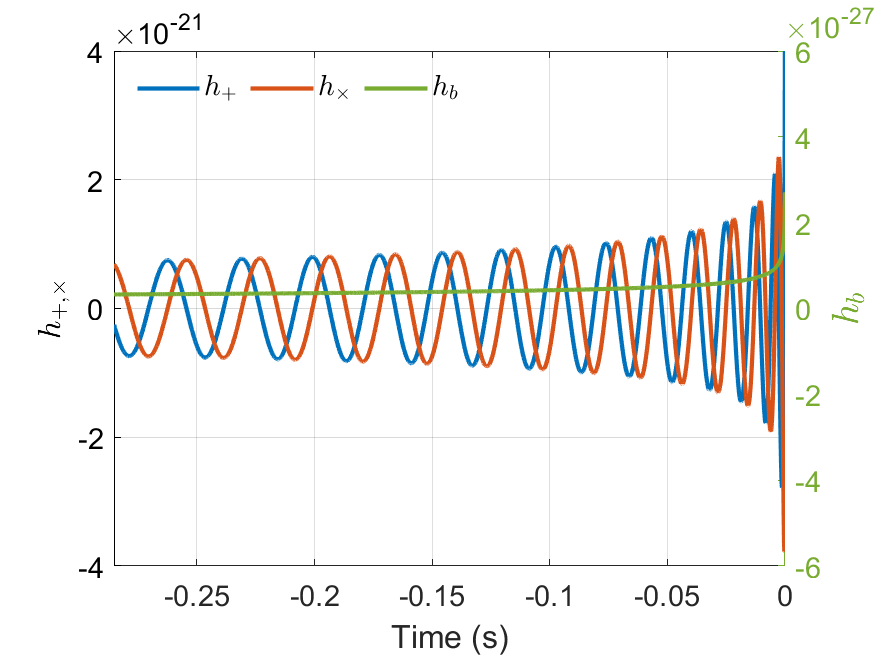}
    & 
    \includegraphics[width=0.49 \textwidth]{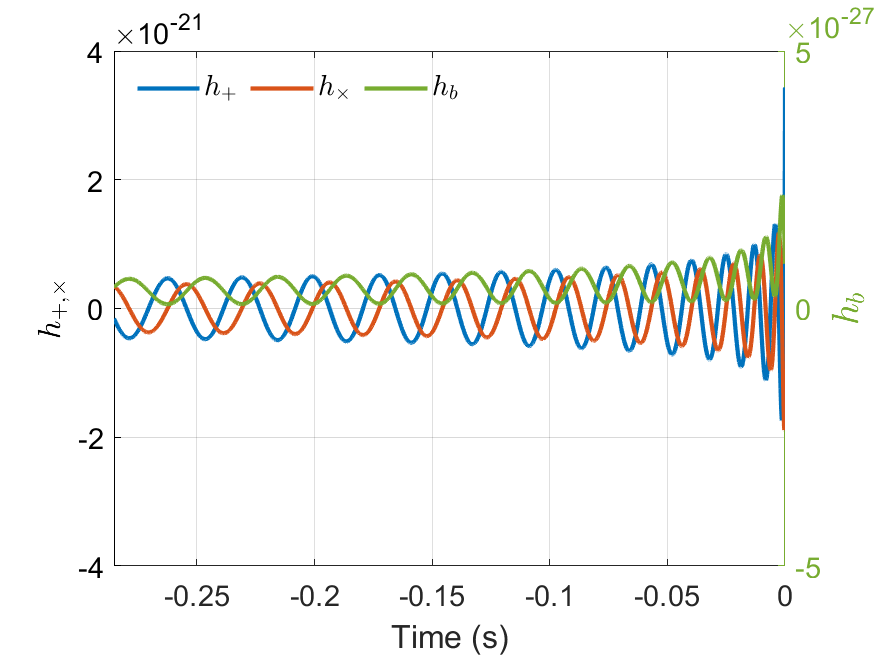}
    \\
    \includegraphics[width=0.49 \textwidth]{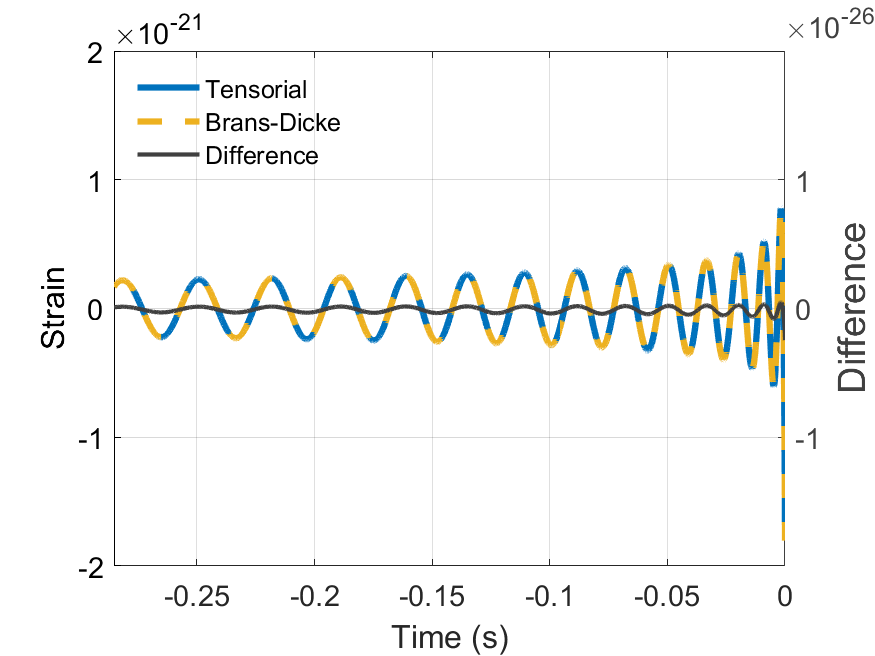}
    & 
    \includegraphics[width=0.49 \textwidth]{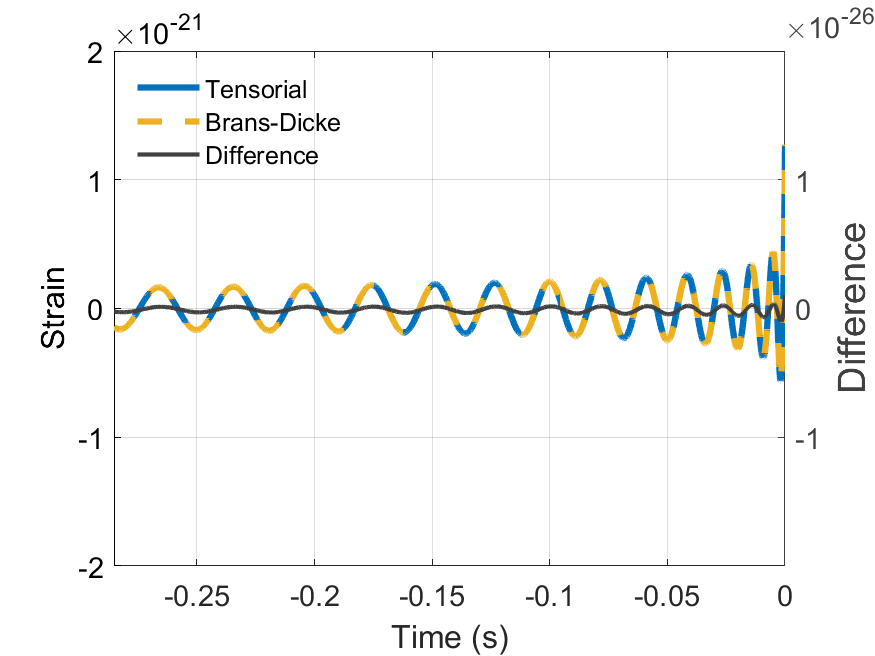}
    \\
    (a) & (b)
    \end{tabular}
    \caption{
    Brans-Dicke GWs polarizations $h_+$, $h_\times$, and $h_b$ (top figures) and the induced strain by tensorial and Brans-Dicke polarizations (bottom figures) for a binary system with parameter values from the event GW150914~\cite{PhysRevLett.116.221101}; and for inclination angles $\iota=0$ (a) and $\iota=\pi/3$ (b).
    The polarization $h_b$ exhibits a very small amplitude, in fact, six orders of magnitude smaller than the ones coming from $h_+$, $h_\times$; and a time-dependant signature which is different from the classical chirp. 
    Moreover, note that the induced strain is rather similar between solely tensorial and Brans-Dicke polarizations, having only minuscule differences due to the small breathing polarization.}
    \label{fig:GWpolarizations_TenAndBransDicke}
\end{figure}

Unlike the non-relativistic case eq.~\eqref{GWstrain}, the strain signal contains three terms: one for each polarization; this is due to the different phases of the cosine functions. Then, fig.~\ref{fig:GWpolarizations_TenAndBransDicke} shows the resulting GWs polarizations from Brans-Dicke theory ($h_+$, $h_\times$ and $h_b$); and the induced strain coming from the only tensorial and Brans-Dicke polarizations for a binary system with parameter values from the event GW150914~\cite{PhysRevLett.116.221101}. Additionally, to illustrate this scenario we include characteristic sensitivities for Brans-Dicke black holes $s1=s2=0.5$, and the coupling constant $\omega_{BD}=10^6$. And, these waveforms are also computed with phase and frequency corrections at the Newtonian approximation. The breathing polarization presents a very low amplitude in comparison to the ones of the $+$ and $\times$ modes; in fact, $h_b$ is six orders of magnitude smaller than $h_+$ and $h_\times$. Moreover, the signature of $h_b$ behaves differently than the typical chirp of the $+$ and $\times$ ones. Particularly, when $\iota=0$, $h_b$ slowly and steadily is increasing up to the coalescence time, whilst it also contains the typical chirp for $\iota=\pi/3$. Due to the tiny breathing polarization mode, the induced strain is rather similar to the solely tensorial polarization, and in fact, it is independent of the inclination angle. The subtle difference is at the order of $10^{-27}$ (see second row of fig.~\ref{fig:GWpolarizations_TenAndBransDicke}).

\subsection{Binary coalescence approximation in Rosen theory}\label{R-theory}

\begin{figure}[t]
    \centering
    \begin{tabular}{cc}
    \includegraphics[width=0.49 \textwidth]{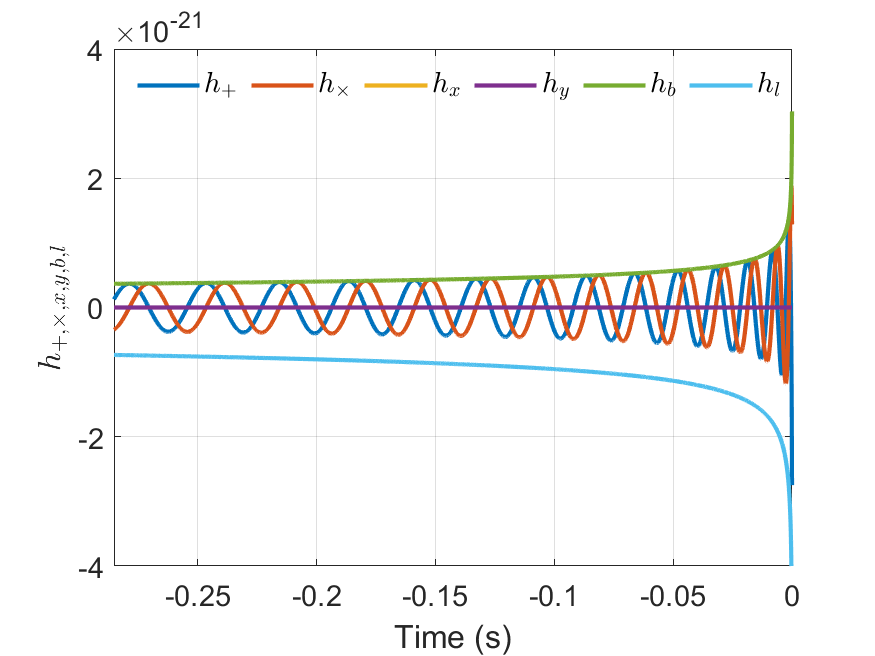}
    & 
    \includegraphics[width=0.49 \textwidth]{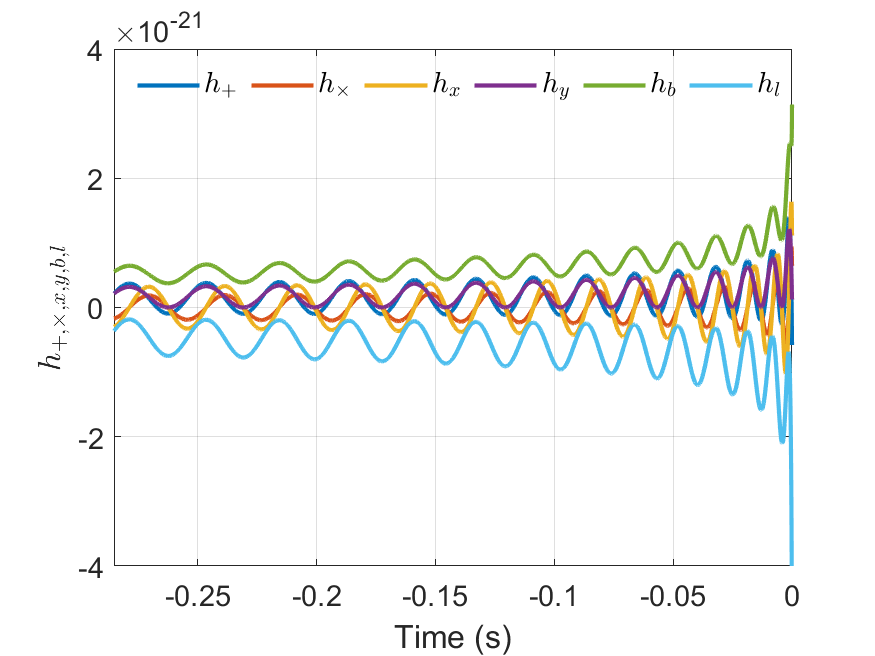}
    \\
    \includegraphics[width=0.49 \textwidth]{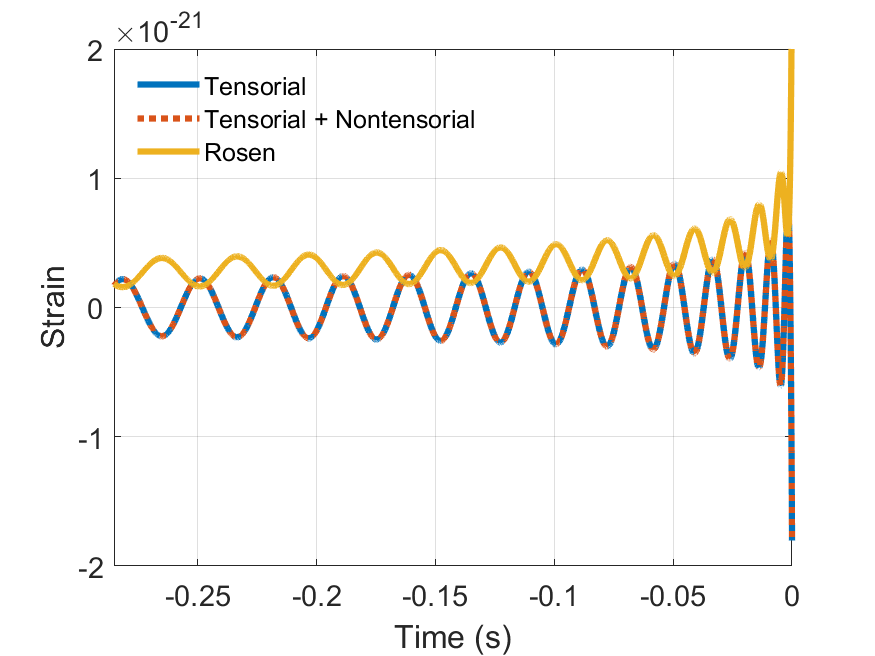}
    & 
    \includegraphics[width=0.49 \textwidth]{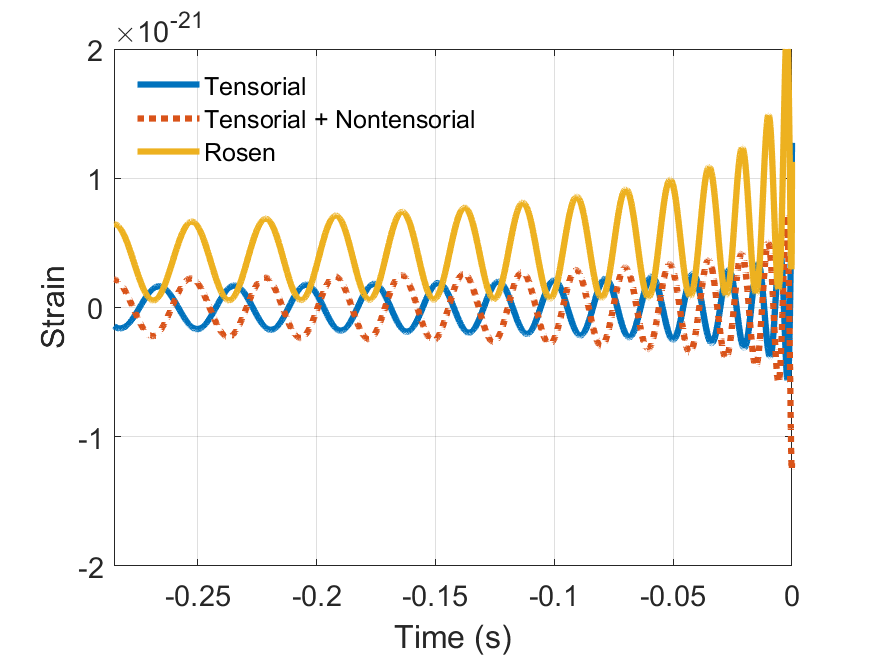}
    \\
    (a) & (b)
    \end{tabular}
    \caption{
    Rosen GWs polarizations $h_+$, $h_\times, h_x, h_y, h_l$ and $h_b$ (top figures) and the induced strain by tensorial and Rosen polarizations (bottom figures) for a binary system with parameter values from the event GW150914~\cite{PhysRevLett.116.221101}; and for inclination angles $\iota=0$ (a) and $\iota=\pi/3$ (b). Note that the Rosen strain signal is a little stronger in amplitude than the tensorial and non-tensorial modes of the Relativistic toy model, when the system is not aligned with the interferometer.}
    \label{fig:GWpolarizations_ER}
\end{figure}

An alternative theory of gravity in which all six polarization modes are present is the Rosen (R) model~\cite{Rosen:1971qp, ROSEN1974455}. This theory represents a bimetric theory of gravity and agrees with GR at the Newtonian limit. It has been proved that this scenario predicts dipole radiation modes; and these might be observed independently of the other modes of polarization. Moreover, it has been tested by analyzing extra-solar system observations, such as the Hulse and Taylor binary pulsar~\cite{Will:1977zz}. Future studies point towards analyzing the behavior of its tensorial and vectorial sectors, in order to determine the GWs response functions and their possible detection. We follow a similar approach from the previous subsection (\ref{GWs-P-BD}), in order to analyze the behavior of the polarization modes present in this theory. The polarization modes for a compact BBH system in terms of chirp mass for Rosen model \cite{Chatziioannou_2012, Will:1977zz}, are represented by: 
\begin{eqnarray}
\label{rosensmodes}
h^{+}&=&\frac{{\cal A}(t)}{2 r }\left [ \sin^2{\phi(t)}-\cos^2{\iota}\cos^2{\phi(t)} \right ] \,,
\\
h^{\times}&=&-\frac{{\cal A}(t)}{2 r }\sin{2\phi(t)}\cos{\iota}\,,  \\
h^{x}&=&\frac{{\cal A}(t)}{r} \left[-\cos{\phi(t)}\sin{\phi(t)}\sin{\iota}-\frac{2}{3}\left(\frac{r}{m}\right)^{1/2}\mathcal{G_B}\sin{\phi(t)}\right] \,,
 \\
h^{y}&=&\frac{{\cal A}(t)}{r}\left[\sin^2{\phi(t)}\sin{\iota}\cos{\iota}-\frac{2}{3}\left(\frac{r}{m}\right)^{1/2}\mathcal{G_B}\cos{\phi(t)}\sin{\iota}\right] \,,
 \\
h^{b}&=&\frac{{\cal A}(t)}{2 r }\left[\sin^2 \iota \sin^2{\phi(t)}+1-\frac{4}{3}\left(\frac{r}{m}\right)^{1/2}\mathcal{G_B}\sin{\iota}\cos{\phi(t)}\right] \,,
 \\
h^{l}&=&\frac{{\cal A}(t)}{r}\left[\sin^2{\iota}\sin^2{\phi(t)}-1-\frac{2}{3}\left(\frac{r}{m}\right)^{1/2}\mathcal{G_B}\sin{\iota}\cos{\phi(t)}\right] \,,
\end{eqnarray}
where ${\cal G_B}=s_1/m_1 - s_2/m_2$, the amplitude ${\cal A}(t)$ term is equal to eq.~\eqref{amplitud}. On the other hand, the frequency for Kepler's law is:
\begin{equation}
2 \pi f = \left[\left(1- \frac{4 s_1 s_2}{3}\right) \frac{m}{r^3}\right]^\frac{1}{2} \,.
\end{equation}
Then, the strain signal given by eq.~\eqref{strain}, in terms of phase and frequency, is expressed by:
\begin{eqnarray}
h(t)&=&\frac{{\cal A}(t)}{r} \left\{- \left[\frac{F_+}{4} \left( 1 + \cos^2 \iota \right) + \left( \frac{F_b}{2} + F_l \right) \sin^2 \iota + F_y \sin \iota \cos  \iota \right] \cos 2 \phi(t) -\frac{1}{2} \left( F_\times \cos \iota + F_x \sin \iota \right) \sin 2 \phi(t) \nonumber \right. \\
&-& \left. \frac{2}{3} {\cal G_B} \sin \iota \left(\frac{r}{m}\right)^{1/2} \left(F_b+F_l+F_y\right) \cos \phi(t) - \frac{2}{3} {\cal G_B} \left(\frac{r}{m}\right)^{1/2} F_x \sin \phi(t) 
\nonumber \right. \\
&+& \left. F_b \left(-1+ \frac{\sin ^2 \iota}{2}\right) - F_l \cos^2 \iota  + \frac{F_+}{4} \left( 1 + \cos^2 \iota \right)  + \frac{F_y}{2} \sin \iota \cos \iota  \right\} \,,
\end{eqnarray}
after a few algebraic simplifications the strain signal becomes:
\begin{eqnarray}
h(t) &=& \frac{{\cal A}(t)}{r}  \left [ \sqrt{ B_{R}^2 + C_{R}^2} \, \cos (2\phi(t) - \Psi) + \sqrt{ E_{R}^2 + F_{R}^2} \, \cos (\phi(t) - \Psi)+ G_{R}\right ] \,,
\end{eqnarray}
where
\begin{eqnarray}
B_{R} &=& -\left[\frac{F_+}{4} \left( 1 + \cos^2 \iota \right) + \left( \frac{F_b}{2} + F_l \right) \sin^2 \iota + F_y \sin \iota \cos  \iota \right] \,, \\
C_{R} &=& -\frac{1}{2} \left[ F_\times \cos \iota + F_x \sin \iota \right], \\
E_{R} &=& -\frac{2}{3} {\cal G_B} \sin \iota \left(\frac{r}{m}\right)^{1/2} \left(F_b+F_l+F_y\right) \,,\\
F_{R} &=& -\frac{2}{3} {\cal G_B} \left(\frac{r}{m}\right)^{1/2} F_x \,, \\
G_{R} &=& F_b \left(-1+ \frac{\sin ^2 \iota}{2}\right) - F_l \cos^2 \iota  + \frac{F_+}{4} \left( 3 + \cos^2 \iota \right)  + \frac{F_y}{2} \sin \iota \cos \iota \,.
\end{eqnarray}
We define
\begin{eqnarray}
{\cal D}_{R1}&=& \frac{r}{ \sqrt{ B_{R}^2 + C_{R}^2}} \,,  \qquad {\cal D}_{R2}= \frac{r}{ \sqrt{ E_{R}^2 + F_{R}^2}} \,, \qquad \qquad {\cal D}_{R3}= \frac{r}{G_{R}} \,,
\end{eqnarray}
therefore we can express the strain signal by:
\begin{eqnarray}
h(t) &=& {\cal A}(t)  \left [ \frac{1}{{\cal D}_{R1}} \, \cos (2\phi(t) - \Psi) + \frac{1}{{\cal D}_{R2}} \, \cos (\phi(t) - \Psi)+ \frac{1}{{\cal D}_{R3}}\right] \,.
\end{eqnarray}
Fig.~\ref{fig:GWpolarizations_ER} shows the GWs polarizations resulting from the Rosen theory for a binary system with parameter values from the event GW150914~\cite{PhysRevLett.116.221101}. Top figures indicate the behaviour of $h_+$, $h_\times \,, h_x \,, h_y \,, h_l$, and $h_b$; bottom panels present the strain induced $h(t)$. Both sets evolve with respect to time; and plotted for inclination angles $\iota=0$ (a) and $\iota=\pi/3$ (b). Observe that all given polarizations and strain signals lie within the range of $10^{-21}$. Then at $\iota=0$ the modes $h_+$ and $h_\times$ exhibit similar waveforms, contrary to the remaining non-tensorial ones. Also, when $\iota=\pi/3$ all the polarization are oscillatory having different starting point, amplitudes, and phases~\cite{Hilborn:2021}. Indeed, this behaviour becomes important since one could differentiate the signals coming from relativistic and alternatives theories, thus establishing similarities and differences between the models. 

%

\subsection{Binary coalescence approximation in Lightman-Lee theory}\label{LL-theory}

\begin{figure}[t]
    \centering
    \begin{tabular}{cc}
    \includegraphics[width=0.49 \textwidth]{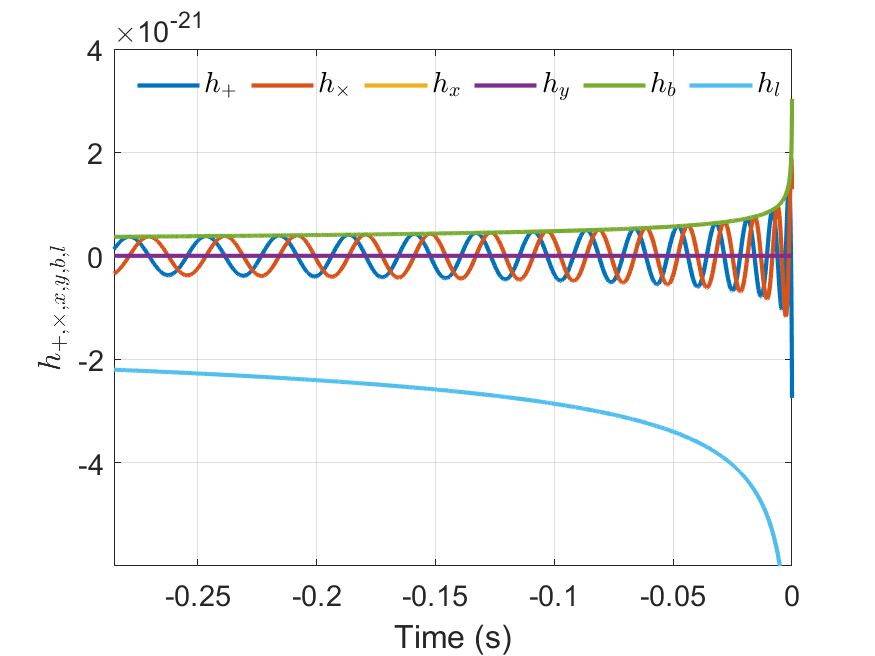}
    & 
    \includegraphics[width=0.49 \textwidth]{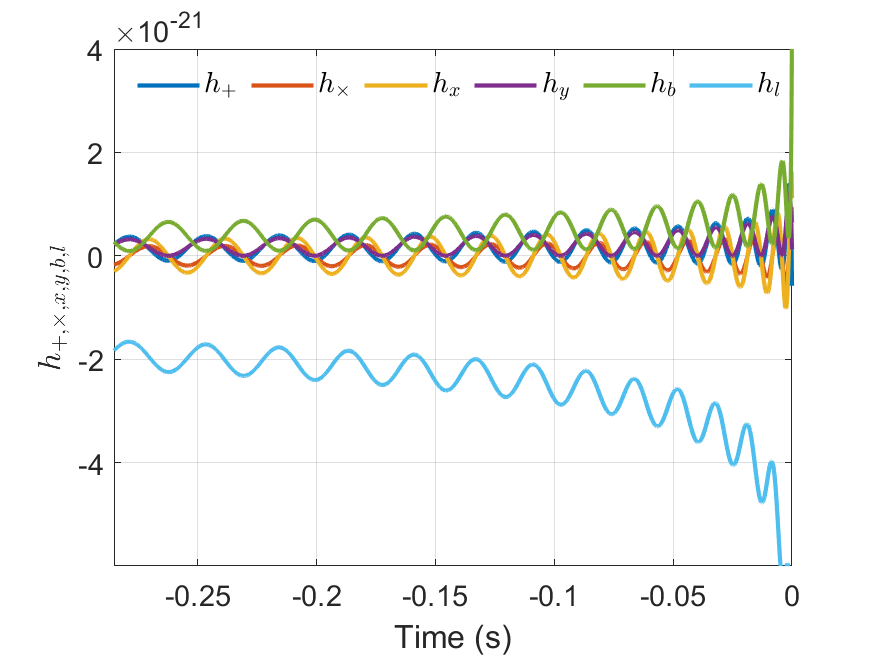}
    \\
    \includegraphics[width=0.49 \textwidth]{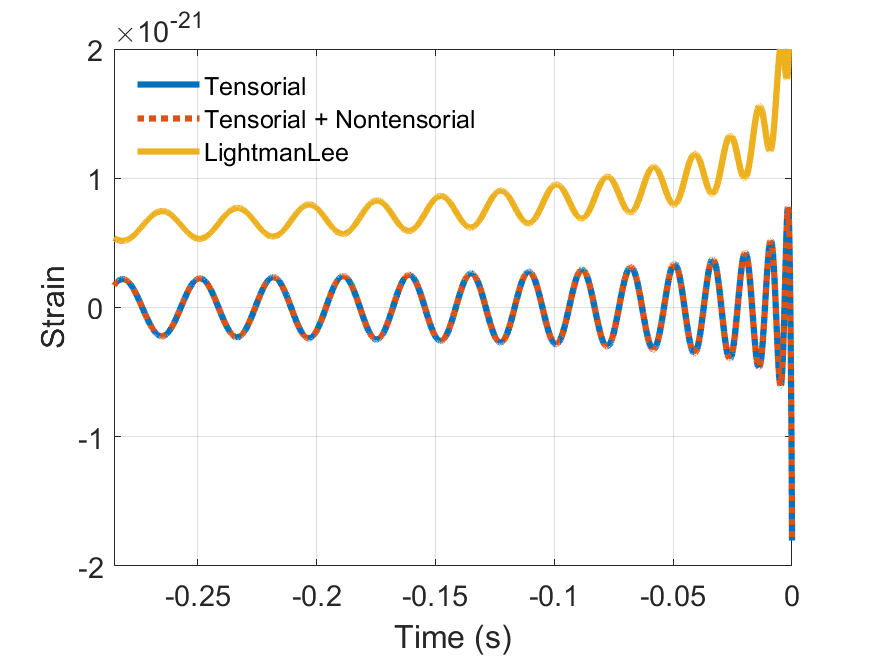}
    & 
    \includegraphics[width=0.49 \textwidth]{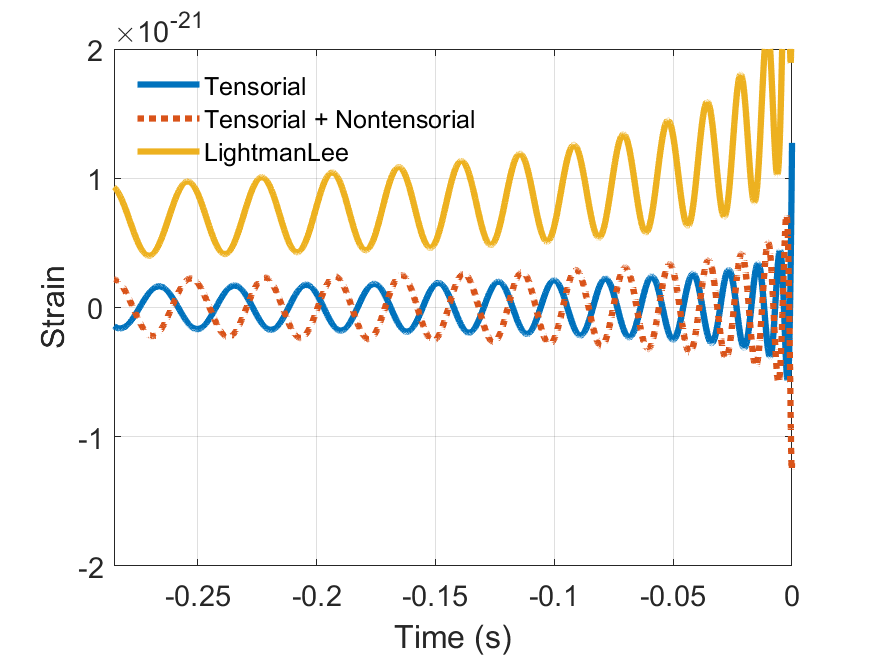}
    \\
    (a) & (b)
    \end{tabular}
    \caption{
    Lightman-Lee GWs polarizations $h_+$, $h_\times, h_x, h_y, h_l$ and $h_b$ (top figures) and the induced strain by tensorial and Lightman-Lee polarizations (bottom figures) for a binary system with parameter values from the event GW150914~\cite{PhysRevLett.116.221101}; and for inclination angles $\iota=0$ (a) and $\iota=\pi/3$ (b). Note that the Lightman-Lee strain signal is a little stronger in amplitude than the tensorial and non-tensorial modes of the Relativistic toy model, when the system is not aligned with the interferometer.}
    \label{fig:GWpolarizations_LL}
\end{figure}

In this section we will obtain the polarization components and the strain signal within the Lightman-Lee (LL) scheme~\cite{PhysRevD.8.3293}. The LL is a bimetric model that has been developed in the Newtonian approximation, and it agrees in this limit with GR theory. This scenario also is consistent with solar-system experiments, and it has been used to determine physical cosmological parameters of alternative theories of gravity. The six polarization modes \cite{Chatziioannou_2012, Will:1977zz, Will:1994fb} are represented by:
\begin{align}
\label{llmodes}
h^{+}&=\frac{{\cal A}(t)}{2 r} \left [\sin^2{\phi(t)}-\cos^2{\iota}\cos^2{\phi(t)} \right] \,,\\
h^{\times}&=-\frac{{\cal A}(t)}{2 r}\sin{2\phi(t)}\cos{\iota} \,, \\
h^{x}&=\frac{{\cal A}(t)}{r}\left[-\cos{\phi(t)}\sin{\phi(t)}\sin{\iota}+\frac{5}{3}\left(\frac{r}{m}\right)^{1/2}\mathcal{G_B}\sin{\phi(t)}\right] \,,\\
h^{y}&=\frac{{\cal A}(t)}{r}\left[\sin^2{\phi(t)}\sin{\iota}\cos{\iota}+\frac{5}{3}\left(\frac{r}{m}\right)^{1/2}\mathcal{G_B}\cos{\phi(t)}\sin{\iota}\right] \,,\\
h^{b}&=\frac{{\cal A}(t)}{2 r}\left[\sin^2{\iota}\cos^2{\phi(t)}-\sin^2{\iota}\sin^2{\phi(t)}+1 -\frac{25}{6}\left(\frac{r}{m}\right)^{1/2}\mathcal{G_B}\sin{\iota}\cos{\phi(t)}\right] \,, \\
h^{l}&=\frac{{\cal A}(t)}{r}\left[\sin^2{\iota}\sin^2{\phi(t)}-3 -\frac{5}{3}\left(\frac{r}{m}\right)^{1/2}\mathcal{G_B}\sin{\iota}\cos{\phi(t)}\right] \,.
\end{align}
Then, the strain signal, given by eq.~\eqref{strain} in terms of phase and frequency, is: 
\begin{eqnarray}
h(t)&=&\frac{{\cal A}(t)}{r} \left\{- \frac{1}{2} \left[\frac{F_+}{2} \sin^2 \iota + F_y \sin \iota \cos \iota +\left(F_b + F_l \right) \sin^2 \iota \right] \cos 2 \phi(t) -\frac{1}{2} \left[ F_\times \cos \iota + F_x \sin \iota \right] \sin 2 \phi(t) \nonumber \right. \\
&-& \left. \frac{5}{3} {\cal G_B} \sin \iota \left(\frac{r}{m}\right)^{1/2} \left(\frac{5}{2}F_b+F_l-F_y\right) \cos \phi(t)+ \frac{5}{3} {\cal G_B} \left(\frac{r}{m}\right)^{1/2} F_x \sin \phi(t) 
\nonumber \right. \\
&+& \left. F_b \left(1 - \frac{\sin ^2 \iota}{2}\right) - F_l \left( 3- \frac{\sin^2 \iota}{2} \right) + F_y \frac{\sin \iota \cos \iota}{2}+ \frac{F_+}{4} \left(1 + \cos^2 \iota \right) \right\} \,.
\end{eqnarray}
We follow a similar prescription from previous subsection~\ref{R-theory}, therefore the strain signal is represented by:
\begin{eqnarray}
h(t) &=& \frac{{\cal A}(t)}{r}  \left [ \sqrt{ B_{LL}^2 + C_{LL}^2} \, \cos (2\phi(t) - \Psi) + \sqrt{ E_{LL}^2 + F_{LL}^2} \, \cos (\phi(t) - \Psi)+ G_{LL}\right] \,,
\end{eqnarray}
where 
\begin{eqnarray}
B_{LL} &=& - \frac{1}{2} \left[\frac{F_+}{2} \sin^2 \iota + F_y \sin \iota \cos \iota +\left(F_b + F_l \right) \sin^2 \iota \right] \,, \\
C_{LL} &=& -\frac{1}{2} \left( F_\times \cos \iota + F_x \sin \iota \right) \,, \\
E_{LL} &=& -\frac{5}{3} {\cal G_B} \sin \iota \left(\frac{r}{m}\right)^{1/2} \left(\frac{5}{2}F_b+F_l-F_y\right) \,,\\
F_{LL} &=& \frac{5}{3} {\cal G_B} \left(\frac{r}{m}\right)^{1/2} F_x \,, \\
G_{LL} &=& F_b \left(1 - \frac{\sin ^2 \iota}{2}\right) - F_l \left( 3- \frac{\sin^2 \iota}{2} \right) + F_y \frac{\sin \iota \cos \iota}{2}+ \frac{F_+}{4} \left(1 + \cos^2 \iota \right) \,.
\end{eqnarray}
We define
\begin{eqnarray}
{\cal D}_{LL1}&=& \frac{r}{ \sqrt{ B_{LL}^2 + C_{LL}^2}} \,,  \qquad \qquad {\cal D}_{LL2}= \frac{r}{ \sqrt{ E_{LL}^2 + F_{LL}^2}} \,,  \qquad \qquad {\cal D}_{LL3}= \frac{r}{G_{LL}} \,.
\label{distance}
\end{eqnarray}
Thus, we can express the strain signal as:
\begin{eqnarray}
h(t) &=& {\cal A}(t)  \left [ \frac{1}{{\cal D}_{LL1}} \, \cos (2\phi(t) - \Psi) + \frac{1}{{\cal D}_{LL2}} \, \cos (\phi(t) - \Psi)+ \frac{1}{{\cal D}_{LL3}}\right] \,.
\end{eqnarray}
Fig.~\ref{fig:GWpolarizations_LL} shows the GWs polarizations resulting from the LL scheme for a binary system with parameter values from the event GW150914~\cite{PhysRevLett.116.221101}. Again, top figures indicate the behaviour of $h_+$, $h_\times \,, h_x \,, h_y \,, h_l$, and $h_b$; and bottom panels present the strain induced $h(t)$. Both sets evolve with respect to time; and plotted for inclination angles $\iota=0$ (a) and $\iota=\pi/3$ (b). Observe that all given polarizations and strain signals lie within the range of $10^{-21}$. Similarly to previous example, Rosen theory, at $\iota=0$ the modes $h_+$ and $h_\times$ exhibit alike waveforms, contrary to the other non-tensorial components. And, when $\iota=\pi/3$ all the polarization are oscillatory having different starting point, amplitudes, and phases~\cite{Hilborn:2021}; therefore one could differentiate their physical features amongst the proposed models. 
\\ 

\section{CONCLUSIONS}   

Since the first observation of GWs due to BBH (GW150914), the LIGO/VIRGO/KAGRA collaboration has detected more than 90 astrophysical events with expected and unexpected characteristics, and despite that, these are well explained within GR. Nevertheless, the interest on GWs detection concerning alternatives theories continues growing since it allows us to analyze vector and scalar sectors, provided they are avoidable in the GR framework. Thus, it opens a robust mechanism that allows us to establish the existence and validity of these alternative theories.    

Without imposing the TT gauge, in this work we analyze the generation of GWs by a compact binary coalescence system of two BH at the inspiral phase, within the Newtonian approximation in a non-relativistic toy model and classical alternative theories. To compute the GWs waveforms we use the mass and sky position of the GWs detection GW150914. We obtain numerical templates for the tensorial and non-tensorial polarizations at the inspiral phase in terms of the chirp mass, then we utilize this outcome to determine their detection range and characteristic amplitudes of any additional polarization modes (see Figs.~\ref{fig:GWpolarizations1} to~\ref{fig:GWpolarizations_TenAndBransDicke}). Also, we distinguish at which positioning angle $\iota$ the amplitude of the polarization is more intense (see Fig.~\ref{fig:effective distance}); from the longitude and latitude inclinations at the sky map, we determine which are the preferential positions for the scalar, vector, and tensor polarizations. Moreover, we notice that the strain signal, which include the six polarizations, is in fact a function that depends on the amplitude, effective distance, frequency, and phase for all them. Additionally, the analysis done in this research shows how the effective distance factor $1/{\cal D}_{ap}$, having inclinations angles not optimally oriented, might affect the localizations of non-tensorial polarizations when searching for GWs in alternative theories of gravity. Thus, incorporating more detectors around the world opens the possibility to test these supplementary sectors. 

From Figs.~\ref{fig:GWpolarizations1} and~\ref{fig:GWpolarizations_TenAndBransDicke}, we conclude that tensorial polarizations have equal amplitudes for non-relativistic and Brans Dicke models with a value of $10^{-21}\,$m. On the other hand, the strain for the tensorial plus non-tensorial modes slightly varies, for $\iota=\pi/3$, since the vectorial and scalar sectors are present. Note that in the Brans Dicke scheme we obtain one non-tensorial signal: the breathing polarization, which its origin is in fact the scalar field; albeit its resulting amplitude is very small, $10^{-27}\,$m, therefore this upshot does not have an impact on the tensorial sector, and it does not affect the characteristic strain of the GW. Thereby, the scalar mark of the Brans Dicke theory will be outside the observation range. Besides, from Fig.~\ref{fig:GWpolarizations_TenAndBransDicke} note that, independently of the frequency, all amplitudes of the strain of the characteristic noise amplitudes obtained with tensorial, tensorial plus non-tensorial, and Brans Dicke polarizations have almost equal size; except for the breathing instance of Brans Dicke. Secondly, the characteristic strains for the induced case are the same when the source is sub-optimally oriented, therefore we have a threshold at low frequencies. Nonetheless, regarding to the not sub-optimally oriented sample, the non-tensorial sectors modify the characteristic strain at greater frequencies, at almost an order of magnitude.

Supplementary to non-Relativistic toy model and Brans Dicke theory, we analyze the polarization modes and the strain signals for the Rosen and Lightman-Lee theories. From figs.~\ref{fig:GWpolarizations_ER} and~\ref{fig:GWpolarizations_LL} we observe that these parameters are approximately similar to the non-Relativistic case: $10^{-21}\,$m. However, we can observe a distinct strain signal, being only positive; since the vector and scalar sectors produce a different waveform comparing to the tensor one. Besides, within these theories the effective distance becomes very difficult to compute since the components of the non-tensorial cases are separated. Nonetheless, if any polarization could be observed separately, this would shed light on an enhanced detectability analysis given the position of the sky, as shown in fig.~\ref{fig:effective distance}. This brings an exciting outlook at this matter. 

Finally, we can argue that in order to determine the existence of GWs within alternative theories of gravity, a thorough study of tensorial and non-tensor modes is required. Hence, it becomes crucially important to establish the features that will delimit the detection of these polarizations. Accordingly, in our analysis of a binary system we may conclude that the inclination and polarization orientation angles; and the location of the source at the sky are those quantities that allow us to delimit the amplitude and phase of these polarizations. Moreover, once the improved network of detectors, which is in fact under construction, can locate all polarizations and thus determine the exact location of the source, a better characterization of such tensorial and non-tensorial modes is around the corner. Thus, the search for footprints coming from alternative theories of gravity has a prosperous future.

\begin{acknowledgements}
This work was supported by the CONACyT Network Project No. 376127 {\it Sombras, lentes y ondas gravitatorias generadas por objetos compactos astrofísicos}. The authors thank the anonymous reviewer for helping us to improve our paper. A.C.L. acknowledges CONACYT scholarship. C.M. wants to thank PROSNI-UDG support. R.H.J is supported by CONACYT Estancias Posdoctorales por M\'{e}xico, Modalidad 1: Estancia Posdoctoral Acad\'{e}mica. 
\end{acknowledgements} 

\vspace{.5cm}
{\bf Data availability}

The datasets generated during and/or analysed during the current study were implemented in MATLAB$^@$ and are available in the repository [http://gravitationalwaves.mx/archivos.html] (click on option software).

\bibliography{mainN_EPJP}

\end{document}